\documentclass{elsarticle}
\pdfoutput=1
\usepackage{graphicx}
\usepackage{subcaption}
\usepackage{color}
\usepackage{tikz}
\usepackage{newlfont}
\usepackage{multirow}
\usepackage{array}
\usepackage{longtable} 
\usepackage{ifthen}
\usepackage{alltt}
\usepackage{float,rotating}
\usepackage{ifpdf}

\usepackage{tikz}
\usepackage{graphicx}
\usetikzlibrary{positioning}

\usepackage{enumerate}
\usepackage{enumitem}
 \usepackage[text={6.5in,8.5in},centering]{geometry} 
\newcolumntype{C}[1]{>{\centering\let\newline\\\arraybackslash\hspace{0pt}}m{#1}}
\newcolumntype{L}[1]{>{\raggedright\let\newline\\\arraybackslash\hspace{0pt}}m{#1}}
\newcolumntype{R}[1]{>{\raggedleft\let\newline\\\arraybackslash\hspace{0pt}}m{#1}}
\newcommand{\ba}{\begin{array} }
\newcommand{\ea}{\end{array} }
\newcommand{\bae}{\begin{eqnarray}}
\newcommand{\eae}{\end{eqnarray}}
\newcommand{\bea}{\begin{eqnarray*}}
\newcommand{\eea}{\end{eqnarray*}}
\newcommand{\be}{\begin{equation}}
\newcommand{\ee}{\end{equation}}

\newcommand{\pr}{{\bf Proof}~~}

\usepackage{amssymb}
\usepackage{amsthm}
\usepackage{ mathrsfs }
\usepackage{amsmath}
\usepackage[mathcal]{eucal}

\usepackage{graphicx}
\usepackage{adjustbox}

\usepackage{color}
\usepackage[colorlinks]{hyperref}
\usepackage{lscape}
\usepackage{graphicx}
\usepackage{epstopdf}
\usepackage{multicol}
\usepackage{tabularx}
\usepackage{ctable}
\usepackage{booktabs}
\usepackage{setspace}
\usepackage{bm}

\usepackage{graphicx}
\usepackage{subcaption}

\makeatletter

\newenvironment{subsubcaption}
\makeatother

\begin{document}

%
%

\begin{frontmatter}
\title{Dynamics of Social Interactions and Agent Spreading in Social Insects Colonies: Effects of Environmental Events and Spatial Heterogeneity}

\author[add1]{Xisohui Guo}
\author[add2]{Jun Chen}
\author[add2,add3]{Asma Azizi}
\author[add1]{Jennifer Fewell}
\author[add2,add4]{Yun Kang}
\ead{yun.kang@asu.edu}

\address[add1]{School of Life Sciences, Arizona State University, Tempe, AZ 85287, USA}
\address[add2]{Simon A. Levin Mathematical, Computational, and Modeling Sciences Center, School of Human Evolution and Social Change, Arizona State University, Tempe, AZ 85287, USA}
\address[add3]{ Division of Applied Mathematics, Brown University, Providence RI,  02906}
\address[add4]{Sciences and Mathematics Faculty, College of Integrative Sciences and Arts, Arizona State University, Mesa, AZ 85212, USA}

\begin{abstract}

The relationship between division of labor and individuals' spatial behavior in social insect colonies provides a useful context to study how social interactions influence the spreading of agent (which could be information or virus) across distributed agent systems. In social insect colonies, spatial heterogeneity associated with variations of individual task roles, affects social contacts, and thus the way in which agent moves through social contact networks. We used an Agent Based Model (ABM) to mimic three realistic scenarios of agent spreading in social insect colonies. Our model suggests that individuals within a specific task interact more with consequences that agent could potentially spread rapidly within that group, while agent spreads slower between task groups.  Our simulations show a strong linear relationship between the degree of spatial heterogeneity and social contact rates, and that the spreading dynamics of agents follow a modified nonlinear logistic growth model with varied transmission rates  for different scenarios. Our work provides an important insights on the dual-functionality of physical contacts. This dual-functionality is often driven via variations of individual spatial behavior, and can have both inhibiting and facilitating effects on agent transmission rates depending on environment.  The results from our proposed model not only provide important insights on mechanisms that generate spatial heterogeneity, but also deepen our understanding of how social insect colonies balance the benefit and cost of physical contacts on the agents' transmission under varied environmental conditions.

\end{abstract}

\end{frontmatter}

%
%



\section{Introduction}\label{introduction}
Social insect colonies provide one of the most fascinating and tractable contexts for theoretical and empirical explorations of biological complex adaptive systems \cite{wilson19780}. The colonies function as decentralized systems for communications and collective actions \cite{fewell2003social,gordon1996organization}. Lacking a central or hierarchical controller, group-level decisions in the colony are attained primarily via the spread and amplification of information communicated at a local level. Colonies use these self-organizational processes to respond and adapt  to variable environment, to reach consensus when a single decision is required, and to distribute individuals across different roles, as in colony task organization \cite{pratt2005quorum}. In social insect colonies, the role of interactions between nestmates in coordinating group level behavior have been investigated through a diversity of behaviors, including food distribution \cite{cassill1999information}, social defense \cite{hermann1984defensive}, social immunity \cite{hart2001task}, and nest site selection \cite{pratt2005quorum}, as well as more generally in the recruitment of individuals across tasks \cite{gordon1999encounter}. \\

Network models of social insect colonies via local interactions have focused primarily for the spread of three main classes of agents: information, food, and pathogens. In the context of information spreading, colonies should theoretically be organized in a way that allows individuals to transmit relevant information as quickly and accurately as possible \cite{franks1999information}. On the other hand, the rapid development of network contacts is problematic to the control and regulation of pathogen spread through contagious interactions \cite{schmid1998parasites,hart2001task}. Social insects colonies rely on social interactions to balance the need for distribution of useful sources efficiently and the demand to minimize the threat of pathogen spreading by contagious infection through interactions. Kappeler et al. \cite{kappeler2015sociality} show that  the division of a network into subgroups with higher connectivity can inhibit the initial spreading of contagion through social networks, while may rapidly increase spreading within network subgroups or clusters. Task fidelity, shown as a stable and individual pattern of spatial occupancy age-induced, could contribute to this network structure \cite{mersch2013tracking}. Thus, it is important to understand the connection between individual interactions in relation to work demand, and their influences on the spreading of other information (or pathogens) through the entire colony.\\

Workers of social insect colonies differentially distribute themselves across colony space in part based on their task roles \cite{tschinkel2017vertical}, generating Spatial Fidelity Zones (SFZs) in the nest, see Figure (\ref{fig:tasl-loc}) \cite{sendova1995spatial}. This spatial based structure can help regulating local contact rates \cite{gordon1993function}, shaping the structure of networks \cite{mersch2013tracking}, and enhancing communication efficiency for task performance \cite{sendova1994social}. Nevertheless, individual workers do not adhere to strict spatial rules. For example, in Temnothorax rugatulus colonies, ants within a task group have varying time budgets for movement through the nest \cite{charbonneau2015lazy}, and temporary changes in individual movement patterns can enhance or reduce SFZs.  Differential movement patterns among workers have also been shown to influence information flow in the contexts of alarm signal transmission \cite{regnier1968alarm,wilson1971evolution}, and food distribution \cite{sendova2010emergency}.\\

Mathematical models have been an important tool to understand how spatial and environmental effects on social contact dynamics and agent-spreading dynamics through the colony.  {Both information and contagious disease  are spreading through physical contacts in social insect colonies \cite{lu2011small}. Thus, the flow of information has been studied under the framework of innovation diffusion \cite{coleman1966medical} and epidemic infection \cite{daley1964epidemics}.  Gernata et al. \cite{gernat2018automated} simulated spreading-agents via an SI model in an empirical trophallaxis network and explored general  similarity between communication network of human and social insects, despite of difference in speed of their  spreading dynamic. The spreading dynamic in social insects is much faster than in humans even after breaking edges in their social network \cite{gernat2018automated}. An SIS-structured model for the spread of information was developed by \cite{richardson2017short} to investigate the influence of activity cycles on information spread through social insect colonies. Through simulations, they found out that  short-term activity cycles on dynamic time-ordered  contact networks inhibit transmission of  information.
 There are some research focusing on the effects of spatial structure on dynamics of disease/host \cite{mollison1985spatial,barlow1991spatially,barlow2000non,hassell1991spatial}  and the dynamic diffusion rate of information \cite{strang1993spatial}, we still have little understandings of mechanisms that generate spatial heterogeneity and 
 how individual moving preferences affects social contact dynamics and agent's spreading at different environment. In addition, 
 there is a need for us to understand  how social insect colonies, with flexible movement styles, obtain the optimal performances of social networks, such as facilitating the spread of useful information and resources, but restricting the transmission of the harmful information and substances, like poisons and pathogens \cite{onnela2007structure,romano2010social,richardson2015beyond}}\\

 Social insect colony is a great biological system that allows us to use agent based models to explore how spatial organization and local interactions affect information flow through contact networks \cite{blonder2011time}. 
In this work, we propose and study a discrete-time Markov chain model to explore spatial and environmental effects on social contact dynamics and spreading dynamics of agent such as information, pathogen. Our proposed agent-based social interaction dynamical  model incorporates varied task groups and individual spatial walking preference in relation to the assigned task group. To mimic the realistic transition of agent initial spatial distributions  corresponding to three different environmental events, we vary agents initial spatial distribution from random-mixing to aggregated one.  We then quantify the process of information propagation under different initial spatial distributions of social insects workers. {The nonoscillatory information spread process of our model is similar to the  individual-based predator–prey model proposed in \cite{pascual1999individuals}  aims to identify at what level spatial factors can impact the propagation of information through a mean-field approach. } In our model simulations, we monitor dynamical interactive behavior of workers and information transmission in multiple scenarios. We further estimated the agent propagation rate  over the colony from the first seed in the modified logistic regression model. We also apply an estimator of clumping to quantify social insects heterogeneous distribution, and examine the relations among spatial heterogeneity, interaction and information spread at the colony level.\\

\section{Method}\label{method}

We use  an agent-based discrete-time Markov chain model to model a $K\times K$ grid colony of $N(\leq K^2)$ workers of social insect colonies  as  set of anonymous agents.
 Each grid, occupied by at most one worker,  captures spatio-temporal dynamics resembling the  real system. 
 At any given time t worker $\mathscr{A}$ is charactrized by its attribute $\eta_t(\mathscr{A})=(l_t(\mathscr{A}), p_t(\mathscr{A}),w_t(\mathscr{A}), f_t(\mathscr{A}))$, where $l_t(\mathscr{A})$ is the  location of  worker $\mathscr{A}$,  $p_t(\mathscr{A})$ is its task, $w_t(\mathscr{A})$ is its walking style, and $f_t(\mathscr{A}))$ is its information -\textcolor{black}{or pathogen, here we use information as one of cases of spreading agents}- status at time t. Now we explain each component of the attribute $\eta$ separately:
 
 \noindent \underline{\textit{Location and neighboring}}:  \sloppy Worker $\mathscr{A}$ at time t takes at most one of the grid cells in the colony $X = \left\{ \left( i,j \right): 1\le i\le K,1\le j\le K \right\} $, that is, $l_t(\mathscr{A})=l=(i,j)$ such that $1\le i\le K,1\le j\le K$. 
 Workers do not necessarily  know their own inner state. Naturally,  workers  sensing mainly depends on antennation and tactile sensation. The use of visual signals in  workers  is very minor \cite{holldobler1990ants}, and it is  unlikely for them to perceive neighbors more than $1.2cm$ away \cite{gordon1993function}. With the assumption that   workers can sense and interact with their neighbors within the $1$ lattice (1 cm),
we define the set of neighboring cells   as the cells  in the above, below,  right, or left of worker $\mathscr{A}$:
\begin{align*}
NC_t(\mathscr{A})=\{(i\pm 1,j), (i,j\pm 1)~ if ~ l_t(\mathscr{A})=(i,j)\}.
\end{align*} 
For the workers on the edge or in the corner of colony, the size of this neighboring cells will reduce to three and two.  Similarly, set of its neighbors at time t is defined as 
\begin{align*}
N_t(\mathscr{A})=\{\mathscr{B}: l_t(\mathscr{B})\in NC_t(\mathscr{A})\}.
\end{align*}  Therefore, for any worker $\mathscr{A}$  at any time t we have $|N_t(\mathscr{A})|\leq 4,$ and if the worker  is on edge of colony or at the corner this maximum number of neighbors will reduce to three or two.
  
\begin{figure}
     \centering
     \includegraphics[width=.45\textwidth]{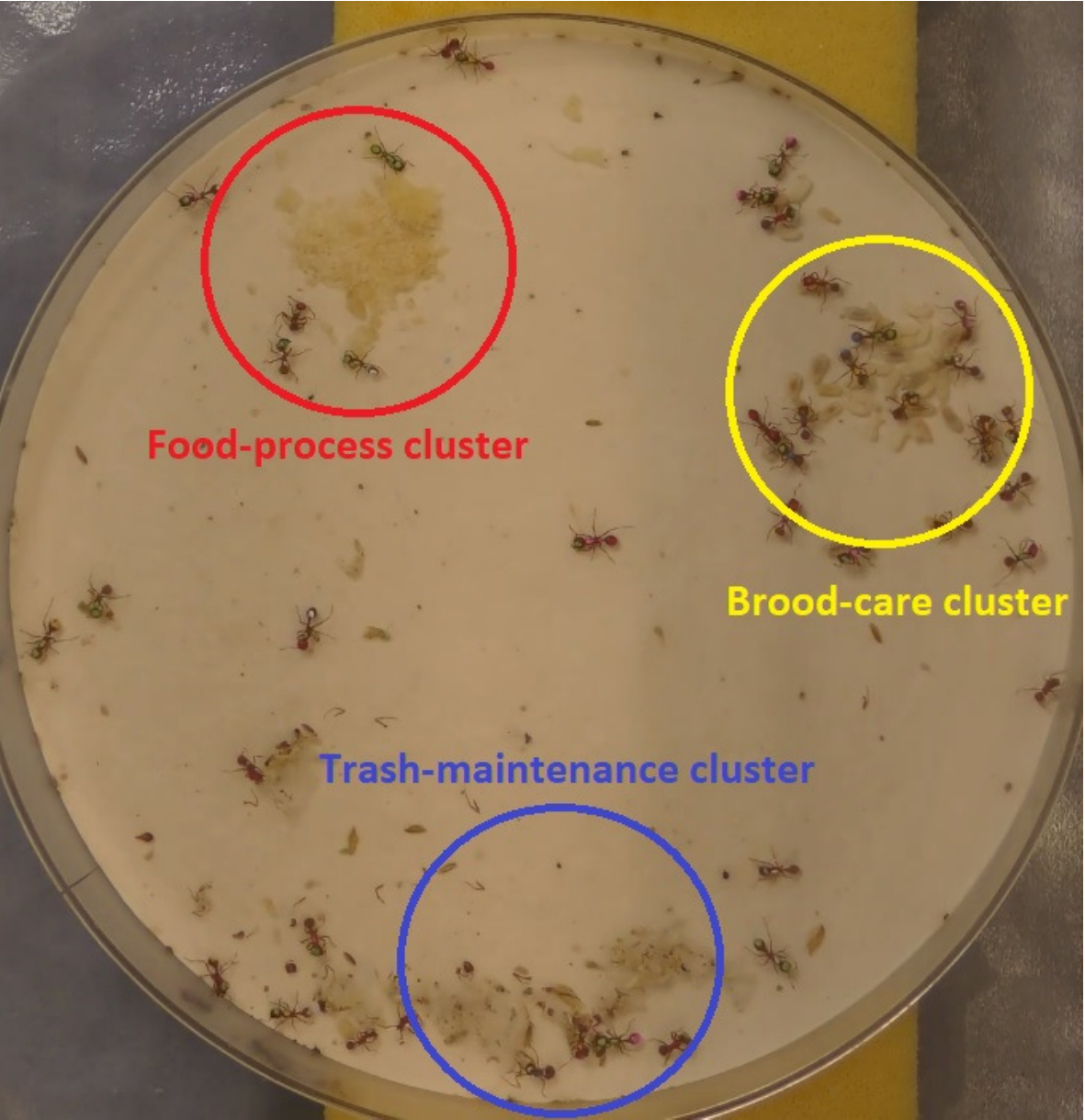}
     \caption{\normalsize{Workers with three different task zones are clustered in their task SFZ as food-processing cluster, brood-care cluster and trash-maintenance cluster. }}
     \label{fig:tasl-loc}
 \end{figure}

 \noindent \underline{\textit{Task group}}: Based on the laboratory observations on the  social insects colonies (\textit{P. californicus}), three major task zones that  workers aggregated around are usually formulated in the colony: brood-care cluster, trash-maintenance cluster, and food-processing cluster Figure \ref{fig:tasl-loc}. There is $P$ different task group that each worker takes exactly one of them at a time. For each task $p\in\{1,2,...,P\}$ we allocate one central location- called SFZ-  in the colony called $S_p\in X$. This SFZ for each task is disjoint from other task, that is, $S_p\neq S_{q}$ if $p\neq q$. The Figure \ref{fig:tasl-loc} shows how workers with different tasks are clustered in locations related to their task, SFZs.
 
 We also assume there is no task switching in the model, that is, worker  $\mathscr{A}$ keeps its initial task  for all the time, $p_t(\mathscr{A})=p_0(\mathscr{A})$ for all $t$s. With that assumption, we can partition  $N$ workers to sum of $N_p$s where $N_p$ is the number of workers with task p:
  $$N=\sum_{p=1}^PN_p=\sum_{p=1}^P|\{\mathscr{A}: p_0(\mathscr{A})=p\}|.$$
 \noindent \underline{\textit{Walking style}}: We have two different walking style for social insects colonies: Random (R), in which worker  $\mathscr{A}$ randomly selects one of the neighboring cells and move toward that, or Drifted (D) in which worker $\mathscr{A}$ has some preferential direction toward  its task SFZ, that is, if $p_t(\mathscr{A})=p$ and  $w_t(\mathscr{A})=D$ then $\mathscr{A}$ moves to one of the neighboring cells  closest to $S_p.$ Similar to task, walking style for each worker   is predetermined at time $t=0$ and is fixed for all future time $t>0$.  Therefore, each task group $N_p$  can be divided into two sets: set of workers with task $p$ who perform random walking and the set of workers with the same task who perform drifted walking style. Based on that we define  the  \emph{spatial fidelity} (SF) of the task group $p$:
 	\begin{equation}
	SF(p) = \frac{|\{\mathscr{A}: p_t(\mathscr{A})=p~\textnormal{and}~w_t(\mathscr{A})=D\} |}{N_p},
    \label{sf-equ}
	\end{equation}
	that is the fraction of workers with task $p$ having drifted walking style.\\
  \noindent \underline{\textit{Information}}: Information with a property that can initiate a change in the state of the receiver advertently (a signal) or inadvertently (a cue), could be transmitted in the colony to complement individual decision-making capability on task performances. At time $t$ we  categorize  worker   $\mathscr{A}$ as informed $f_t(\mathscr{A})=1$, or not informed $f_t(\mathscr{A})=0$. An informed worker  can spread information to other not informed neighbor workers with some probability $\beta_i$.\\

 Now we explain the dynamic of movement and  information spread of  social insect colonies through time. 
 We assume each update, i.e.,  one-time tick, is  consistent with $\Delta t$. We also assume that the basic speed of workers is one cell per time step. Workers cannot cross the reflecting walls and borders, instead when they reach the borders and walls, they will redirect randomly. At any time $t$ we select a worker  with attribute $\eta_t(\mathscr{A})=(l_t(\mathscr{A}),p_t(\mathscr{A}),w_t(\mathscr{A}),f_t(\mathscr{A}))=(l,p,w,f)$ from the total population of $N$ workers randomly to move to one of the cells $\in NC_t(\mathscr{A})$ randomly.  If the selected cell is occupied with one of the neighbor worker $\mathscr{B}$ we say  $\mathscr{A}$ and $\mathscr{B}$  have contacts, otherwise  $\mathscr{A}$ performs walk.
At any given time t, each worker $\mathscr{A}$ can change one or all of its attributes through the following procedure:\\
 
\begin{figure}[!ht]
\centering
\includegraphics[width=1\textwidth]{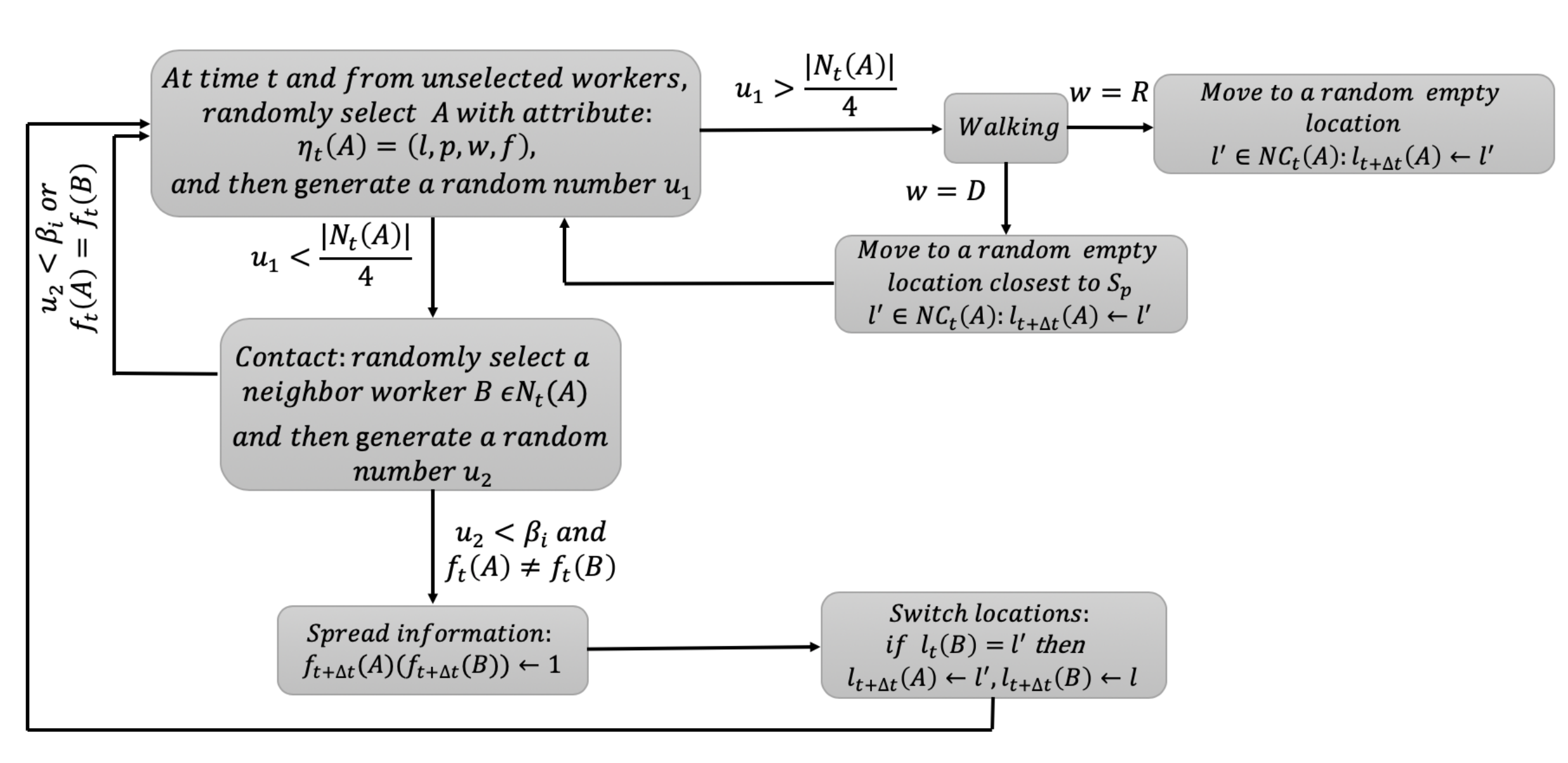}
\caption{\normalsize{Modeling schematic diagram of our social network model.}}
\label{fig:alg}
\end{figure}

 \noindent Randomly select  $\mathscr{A}$ with attribute $\eta_t(\mathscr{A})=(l_t(\mathscr{A}),p_t(\mathscr{A}),w_t(\mathscr{A}),f_t(\mathscr{A}))=(l,p,w,f).$
\begin{enumerate}
    \item  The selected worker  has $|NC_t(\mathscr{A})|$ neighboring cells and $|N_t(\mathscr{A})|$ neighbors, therefore with the probability of $1-\frac{|N_t(\mathscr{A})|}{|NC_t(\mathscr{A})|}$, $\mathscr{A}$ walks into an empty location with the following rules:
    
\begin{enumerate}[label=(\alph*)]
\item If $w=R$, the chosen worker  has a random walking style, the worker  randomly walks into one of the empty locations $l'\in NC_t(\mathscr{A})$ with probability
\begin{eqnarray*}
P(l_{t+\Delta t}(\mathscr{A})=l' & \vert & \textnormal{$l'$ is empty} ~\& ~w_t(\mathscr{A})=R ) \\
&=& \frac{1}{|NC_t(\mathscr{A})|-|N_t(\mathscr{A})|}\times\frac{|NC_t(\mathscr{A})|-|N_t(\mathscr{A})|}{|NC_t(\mathscr{A})|}\\
&=& \frac{1}{|NC_t(\mathscr{A})|}.
\end{eqnarray*}
\item If $w=D$, the chosen  worker  has a preferential walking style, the worker  walks into one of its empty neighborhood cell $l'\in NC_t(\mathscr{A})$  closest to its task SFZ  $S_p$  \textcolor{black}{(SFZs)}  with probability
\begin{eqnarray*}
P(l_{t+\Delta t}(\mathscr{A})=l' & \vert & \textnormal{$l'$ is empty} ~\& ~,w_t(\mathscr{A})=D )\\&=&\frac{|NC_t(\mathscr{A})|-|N_t(\mathscr{A})|}{|NC_t(\mathscr{A})|}.\end{eqnarray*}
\end{enumerate}
\item  The selected worker has $N_t(\mathscr{A})$ neighbors and therefore, it has a contact with one of its neighbors with probability  $\frac{|N_t(\mathscr{A})|}{|NC_t(\mathscr{A})|}$. 
Assume that the chosen neighbor $\mathscr{B}$ has attribute   $\eta_t(\mathscr{B})=(l_t(\mathscr{B}),p_t(\mathscr{B}),w_t(\mathscr{B}),f_t(\mathscr{B}))=(l',p',w',f').$ We have two cases:
\begin{enumerate}[label=(\alph*)]
\item If $f=f'$  then the two workers switches their location with the following probability:
\begin{eqnarray*}
P(l_{t+\Delta t}(\mathscr{A})=l'~\&~l_{t+\Delta t}(\mathscr{B})=l &\vert & l_{ t}(\mathscr{A})=l~\&~l_{t}(\mathscr{B})=l') \\
&=& \frac{|N_t(\mathscr{A})|}{|NC_t(\mathscr{A})|}.
\end{eqnarray*}
\item If  $f\neq f'$ and without loss of generality we assume $f=1$, that is $\mathscr{A}$ is informed, then the informed worker spreads information to the other one with probability 
\begin{eqnarray*}
P(f_{t+\Delta t}(\mathscr{A})=1~\&~f_{t+\Delta t}(\mathscr{B})=1 & \vert & f_{ t}(\mathscr{A})=1~\&~f_{t}(\mathscr{B})=0) \\
&=& \frac{\beta_i|N_t(\mathscr{A})|}{|NC_t(\mathscr{A})|}.
\end{eqnarray*}
\end{enumerate}
\end{enumerate}
The schematic diagram of our dynamical model and the related variables are shown in the Figure \ref{fig:alg} and Table \ref{tab:par-def}, respectively.
To further  study  how environment and spatial components affect the dynamics of social interactions and information spread in social insects colonies, we first define some concepts.
Let $C(t)$ be the total number of contacts occurred between workers of social insects colony  in the time interval $(0,t)$, then the contact rate $R(t)=\frac{dC}{dt}$ is approximated  by the number of  contacts during the small time interval $\Delta t$:
$$
R(t) \approx \frac{C(t+\Delta t)-C(t)}{\Delta t}.
$$
Similarly, we  define  $R_{pq}(t)$  as the  contact rate between workers with tasks $p$ and $q$. If $p=q$ then $R_{pp}(t)$ is the contact rate within a task group $p$. We also define  $\bar{R}_w(t)=\frac{\sum_{p=1}^P R_{pp}(t)}{P}$ as the average contact rate within same task group, and $\bar{R}_b(t)=\frac{\sum_{ p,q=1\atop p\ne q}^P R_{pq}(t)}{{P\choose 2}}$ as the average contact rate  between different task groups.

Let $P_{l}$ be the probability that  cell location $l$ being occupied by a worker, then we  define \emph{spatial heterogeneity degree} (SHD) of the colony as
\bae\label{SHD}
SHD(t)&=&\frac{\sum_{l=1}^{K^2} \left(P_l-\frac{N}{K^2}\right)^2}{K^2},
\eae
where $\frac{N}{K^2}$ is the probability that a typical  cell $l$ is  occupied by a worker  when  all workers have a  random walk, that is, when $w(\mathscr{A})=R$ for all $\mathscr{A}$s. 
This definition indicates that the smallest value of $SHD$ is the case  when all workers do symmetric random walk ($SHD_{min}=0$), and the largest value of $SHD$ is the case when workers do not move, that is,  $P_{l}=1$ for all $N$ occupied $l$ locations  by $N$ workers, and $P_{l'}=0$ of the remaining $K^2-N$ empty locations $l'$:
	\begin{center}
	$SHD_{max}=\frac{N(1-\frac{N}{K^2})^2+(K^2-N)(0-\frac{N}{K^2})^2}{K^2}= \frac{N(K^2-N)}{K^4}$.
	\end{center}
For simplicity, we resclae  $SHD$ by converting  $K\times K$ grid colony 
	to  $\frac{K}{m}\times\frac{K}{m}$ patches where each patch has $m\times m$ grids. The parameter $m$ is a  conversion parameter, for example, if we have a $300\times 300$ colony, we re-scale it  by choosing $m=10$  and  the number of patches $\frac{300}{10}\times \frac{300}{10}$, that is, each new patch includes $10\times 10$ cells.  Let $P_{l}(\tau)$ be the ratio of occupied grids by workers to all $m\times m$ grids at patch $l$, then we have $SHD(\tau)$ calculated as follows
\bae\label{SHDt}
SHD(\tau)=\frac{\sum_{l=1}^\frac{N}{m} \left(P_{l}(\tau)-\frac{N}{K^2}\right)^2}{K^2}.
\eae
We define  $I(t)$ as the number of informed workers at time $t$:
$$I(t)=|\{\mathscr{A}: f_t(\mathscr{A})=1\}|,$$
The rate $\frac{dI}{dt}$ is approximated  by the number of information received  during the small time interval $\Delta t$:
$$\frac{dI}{dt}\approx \frac{I(t+\Delta t)-I(t)}{\Delta t}.$$

\begin{table}[htp]
\centering
\begin{tabular}{ llp{9.55cm}}
\toprule[1.5pt]
  & \multicolumn{2}{c}{}\\
  & \textbf{{Parameter}} & \textbf{{Description}}
 \\
  \cmidrule(lr){2-3}\cmidrule(l){2-3}
  \multirow{2}{2.5cm}{} & $K\times K$ & Colony size \\
   \multirow{2}{2.5cm}{Colony Parameters and Variables} &  $P$ & Number of different task \\ 
   \multirow{2}{2.5cm}{} &  $SF(p)$ & Spatial fidelity for task p \\ 
   \multirow{2}{2.5cm}{} &  $S_p$& SFZ for task $p$\\ \multirow{2}{2.5cm}{} & $SHD$ & Spacial heterogeneity degree of colony \\
   \multirow{2}{2.5cm}{} & $m$ & The conversion ratio of space \\
\cmidrule(lr){2-3}\cmidrule(lr){2-3}
 \multirow{2}{2.5cm}{} & $N$ & The total number of workers in social insects colony \\
   \multirow{2}{2.5cm}{} &$N_p$ & The total number of workers with task p  \\ 
   \multirow{2}{2.5cm}{} &  $l_t(\mathscr{A})$ & Location of worker $\mathscr{A}$ at time $t$ \\ 
   \multirow{2}{2.5cm}{} &  $p_t(\mathscr{A})$ & Task of worker  $\mathscr{A}$ at time $t$\\ 
    \multirow{2}{2.5cm}{} &  $w_t(\mathscr{A})$ & Walking style of worker $\mathscr{A}$ at time $t$ \\ 
   \multirow{2}{2.5cm}{Worker Parameters and Variables} &  $f_t(\mathscr{A})$ & Information status of worker $\mathscr{A}$ at time $t$\\ 
     \multirow{2}{2.5cm}{} &  $NC_t(\mathscr{A})$ & Set of neighboring cells  of worker $\mathscr{A}$ at time $t$\\ 
        \multirow{2}{2.5cm}{} &  $N_t(\mathscr{A})$ & Set of neighbors of worker $\mathscr{A}$ at time $t$\\ 
              \multirow{2}{2.5cm}{} &  $C(t)$ & Total number of contacts between workers at  time interval $(0,t)$\\ 
                 \multirow{2}{2.5cm}{} &  $R(t)$ & Contact rate  at time $t$\\ 
                    \multirow{2}{2.5cm}{} &    $\bar{R}_w(t)$ & Average within group contact rate at time $t$\\ 
                       \multirow{2}{2.5cm}{} &  $\bar{R}_b(t)$ & Average between groups contact rate at time $t$\\ 
                          \multirow{2}{2.5cm}{} &  $f_t(\mathscr{A})$ & Information status of worker $\mathscr{A}$ at time $t$\\ 
                            \multirow{2}{2.5cm}{} &  $I(t)$ & Fraction of informed workers at time $t$\\ 
                              \multirow{2}{2.5cm}{} &  $\beta_i$ & Probability of information spread\\ 
\bottomrule[1.5pt]
\end{tabular}
\caption{\normalsize {Parameters, Variables and their definition}} 
   \label{tab:par-def}
\end{table}

We use our model to explore how spatial fidelity  affects the  different average contact rates, spatial heterogeneity degree, and information spread in three different environmental scenarios of social insects colony. Each environmental scenario is characterized by the initial configuration of workers and the spatial fidelity as follows:

\begin{enumerate}
\item \textbf{Random-Mixing (RM)}: in which  all workers are randomly distributed in the colony and all of them are assigned with random walking style, that is $SF(p)=0$ for $p\in\{1,...,P\}$. Workers random walking corresponds to random-mixing in \textit{Temnothorax albipennis} after famine emergency \cite{sendova2010emergency}.

\item  \textbf{Random-Initial-Distribution (RID)}: in which all workers are initially distributed in a random location in the colony but  a fraction $f$  follow drifted walking style, that is $SF(p)=f$ for all $p\in\{1,...,P\}$.

\item \textbf{Aggregated-Initial-Distribution (AID)}: in which workers tend to segregate in their task SFZs  \cite{sendova1993task}, we assign $f_p$ fraction of  workers with task  $p$ having drifted walking style, that is, that is $SF(p)=f_p$.
\end{enumerate}

In the next Section we will  study the dynamics of the  contact rate $R(t)$ and its average, the average spatial heterogeneity degree $SHD$ and the agents spreading  defined in this Section under the above   environmental  scenarios. \\
\section{Result}

In this Section, we  perform our analyses and simulations on three different scenarios explained in the Section \ref{method}: RM for $SF=0$, RID and AID for $SF=20\%-98\%$. We will   provide results on the dynamics of the   contact rate $R(t)$ and the averages $\bar{R}_w(t)$, the spatial heterogeneity degree $SHD(t)$ of the colony and the information spread for  different environment scenarios. Each plot is the average of  $40$ different stochastic simulations seeding the same initial condition,  with the model baseline parameters in Table \ref{tab:par-val}, unless stated otherwise.

\begin{table}[H]
 \centering 
   \begin{tabular}{lccr} 
   \toprule[\heavyrulewidth]\toprule[\heavyrulewidth]
   \textbf{Parameter} & \textbf{Baseline value} & \textbf{Parameter} & \textbf{Baseline value} \\ 
   \midrule
   $K\times K$ & $69\times 69$ & $N$ & $180$ \\
   $P$ & $3$ & $N_p$ & $60$\\
    $m$& $3$ & $\beta_i$ & $1$\\
   \bottomrule[\heavyrulewidth] 
   \end{tabular}
   \caption{\normalsize{Parameters and their baseline values}} 
   \label{tab:par-val}
\end{table}

\subsection{Dynamics of the average contact rates R(t) and spatial heterogeneity degree SHD(t)}

\begin{figure}[ht]
\centering
\includegraphics[width=\textwidth]{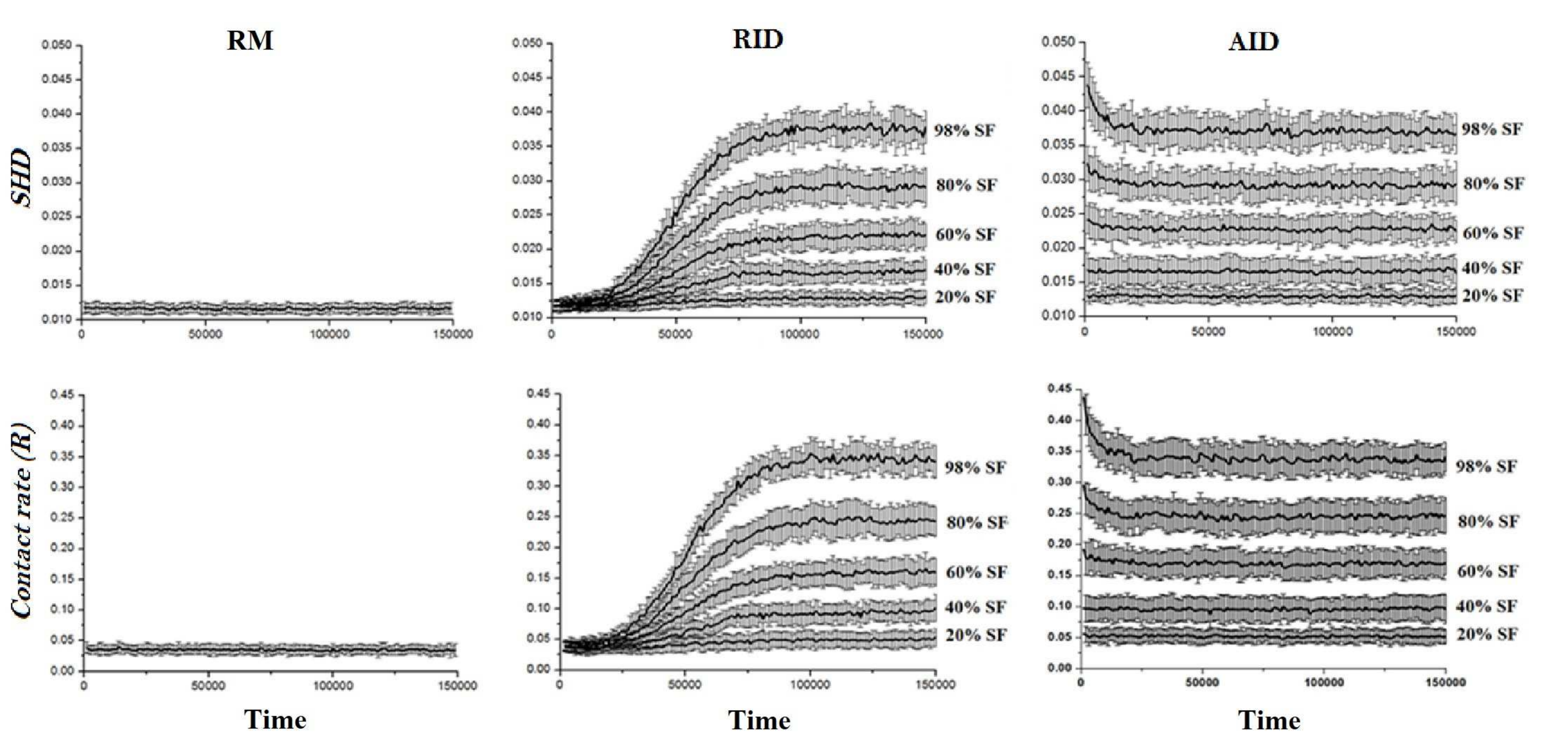}
\caption{ \normalsize{\textbf{Spatial heterogeneity degree  and the average contact rate  over time for RM, RID, and AID scenarios:} Solid black lines represent the average of $40$ replicates on each update and  error bar are $95\%$ confidence intervals. In all scenarios $SHD(t)$ and $R(t)$  share the same trend. $SHD$ synchronizes with the average contact rate of workers in all of the scenarios with various spatial fidelity values.  Biologically, when more workers perform preferential movement the $SHD$ of workers distribution and  their interactive behavior will escalate.}}
\label{fig:spatial}
\end{figure}

In this first subsection we study the dynamic of spatial heterogeneity degree and contact rate and its averages over different environmental scenarios.
The main observation of this part is that both the average contact rate dynamics $R(t)$ and the related spatial heterogeneity degree dynamics $SHD(t)$  follow the logistic growth patterns with different intrinsic growth rates and carrying capacities in the scenarios of RID, Figure \ref{fig:spatial}. 

Spatial heterogeneity degree defined in Equation \ref{SHDt} measures the level of deviation from the even distributions of workers of social insects colonies over the space. 
In the first row of Figure \ref{fig:spatial} we observe that in the RM  scenario  $SHD(t)$ is almost constant over the time with the value of $0.0115$, 
but in  the RID and AID  scenarios dynamics of $SHD(t)$ is not constant but shares similar patterns as its corresponding average contact rate $R(t)$. The Figure \ref{fig:spatial} also  suggests that $SHD$ synchronizes with the average contact rate of  workers in all of the scenarios and for all   spatial fidelity values.  Specifically, more workers perform preferential movement- the higher spatial fidelity $SF$- higher degree of $SHD$ plateau for the  colony and   higher contact rates.

 To further explore the correlation between  $SHD(t)$ and $R(t)$ under different scenarios, we  pooled them pairwise, and observed a linear trend. There is a linear correlation between $SHD(t)$ and $R(t)$ that is  represented by Model Equation \ref{R-SHD}.  \bae\label{R-SHD}R(t)=-0.1033+11.895 SHD(t),
\eae 
with  adj-R$^2=0.9985$,  F-value$=1.10029$, and P-value$<0.001$.
This result means that the social contact network (e.g., the average contact rate) could be formulated by spatial heterogeneity due to  non-random walking styles.
 As consequences, RID could be the scenario interlinking the random distribution of workers initially (RM) and  their  segregation in their corresponding SFZ at the  end (AID).\\

\begin{figure}[!ht]
\centering
\includegraphics[width=0.6\textwidth]{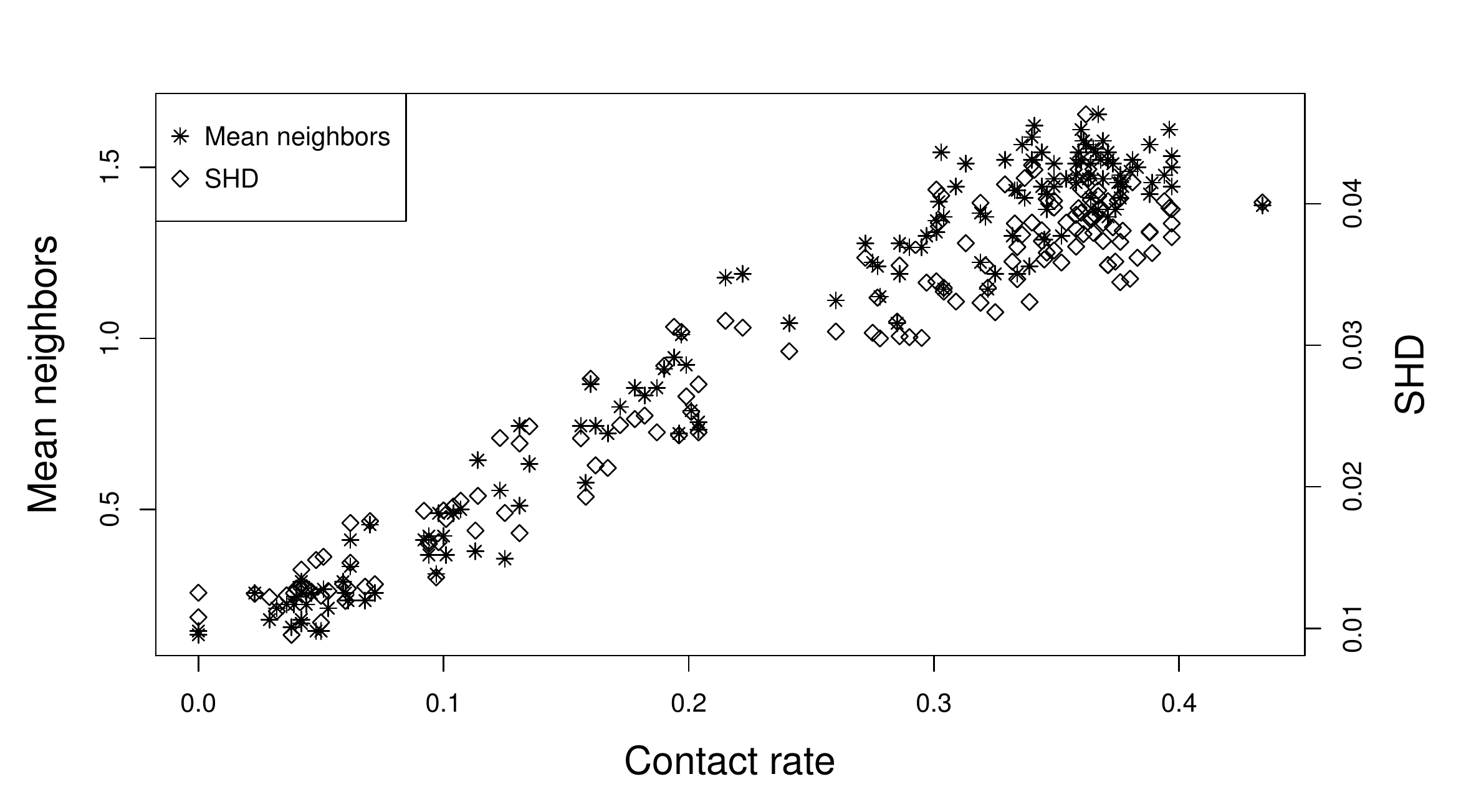}
\caption{\label{fig:overlap}\normalsize{\textbf{Overlapping between mean-neighbor and \pmb{$SHD$} over contact rate in double y-axes for  RID scenario and \pmb{$SF=98\%$}:}there is a  synchronization between  $SHD(t)$ and $R(t)$, that is, that the larger value of $SHD$, the more crowded neighboring space, therefore,  more contact with nest-mates.}}
\end{figure}

Our spatial heterogeneity degree $SHD(t)$ defined in Equation \ref{SHDt} reflects the \textit{"mean-crowding"} concept introduced by Lloyd \cite{lloyd1967mean}. Mean-crowding measures the spatial heterogeneity of the disease/host model, which is calculated by the total number of neighbors every organism has  over the number of organism with at least one neighbor. To illustrate relationships among mean-crowding, $SHD(t)$ and the $R(t)$, we calculated them within the  RID scenario for the spatial fidelity being $SF=98\%$. Both SHD and mean-neighbors increase linearly as the average contact rate increases, Figure \ref{fig:overlap}. This result illustrates  that $SHD$ provides a quantified measure of spatial heterogeneity as the "mean-crowding" concept. Also, the overlapping between $SHD$ and mean-neighbors  offers an explanation for  the synchronization of $SHD(t)$ and $R(t)$ that the larger value of $SHD$ represents the more crowded neighboring space, as a consequence, ensures more opportunities to contact with nest-mates.\\

\begin{figure}[htp]
        \centering
        \begin{subfigure}[b]{0.45\textwidth}
            \centering
            \includegraphics[width=\textwidth]{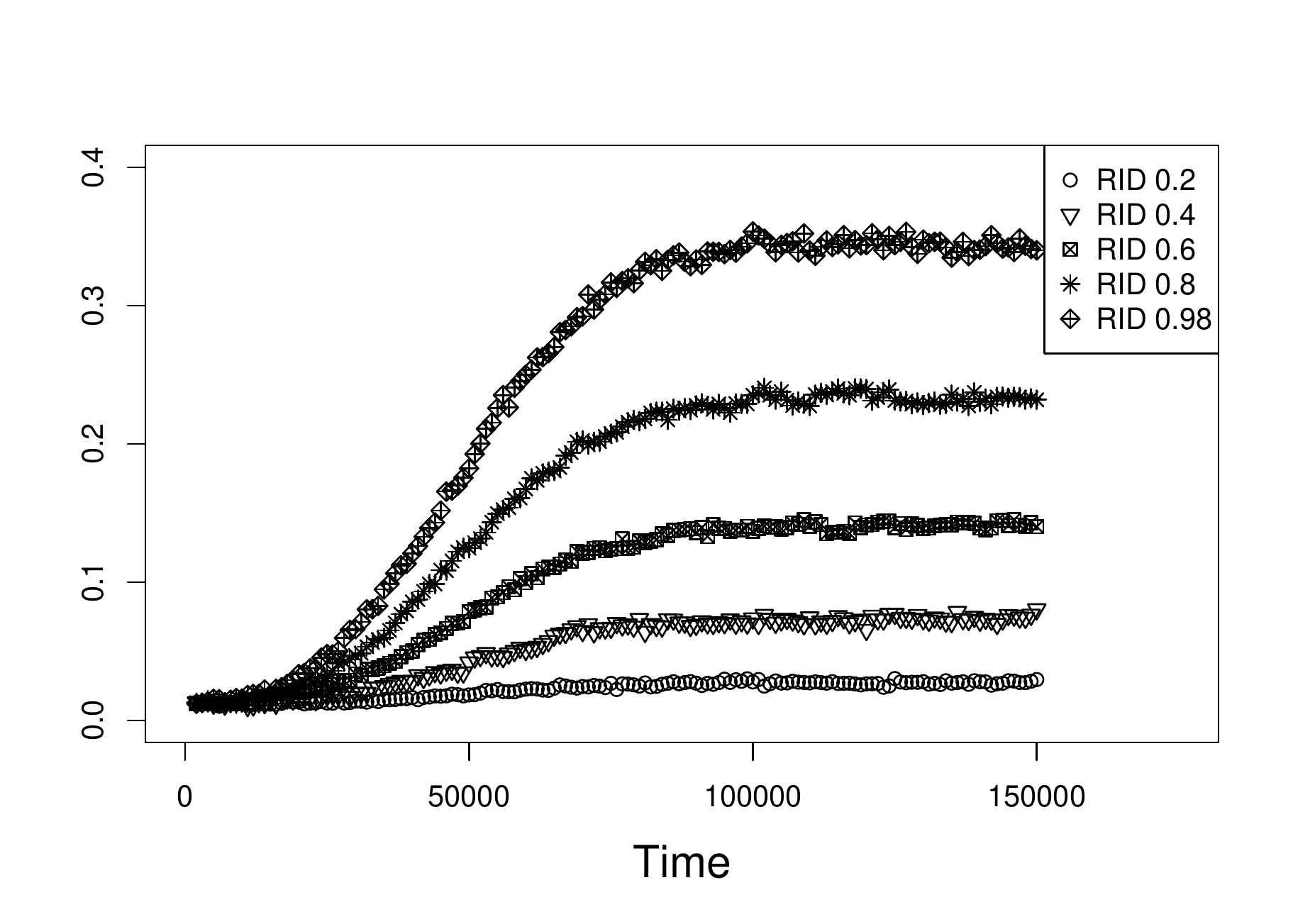}
            \caption{{\small $\bar{R}_w$ for RID Scenario}}    
        \end{subfigure}
        \begin{subfigure}[b]{0.45\textwidth}  
            \centering 
            \includegraphics[width=\textwidth]{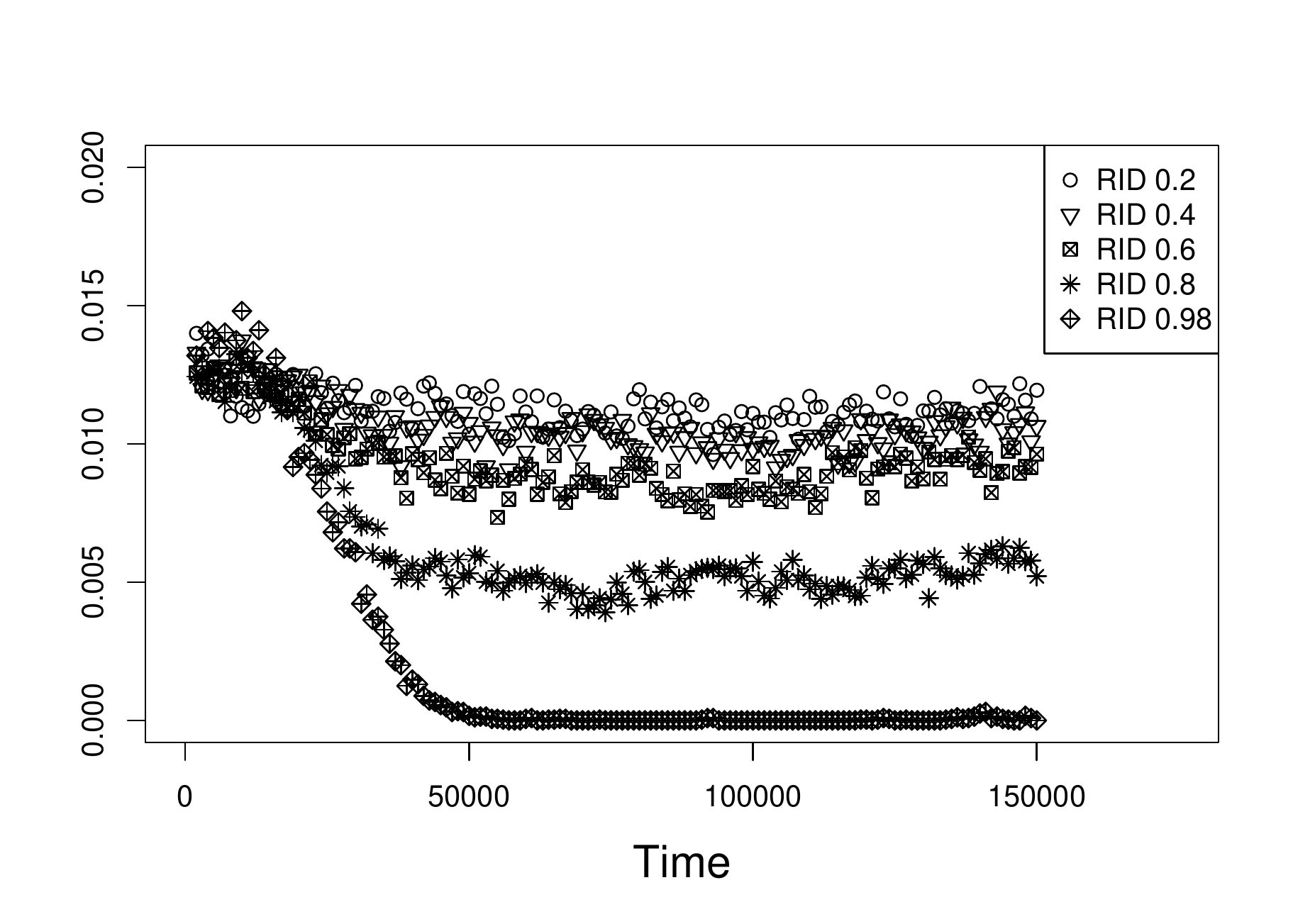}
            \caption{{\small $\bar{R}_b$ for RID Scenario}}   
        \end{subfigure}
        \begin{subfigure}[b]{0.45\textwidth}   
            \centering 
            \includegraphics[width=\textwidth]{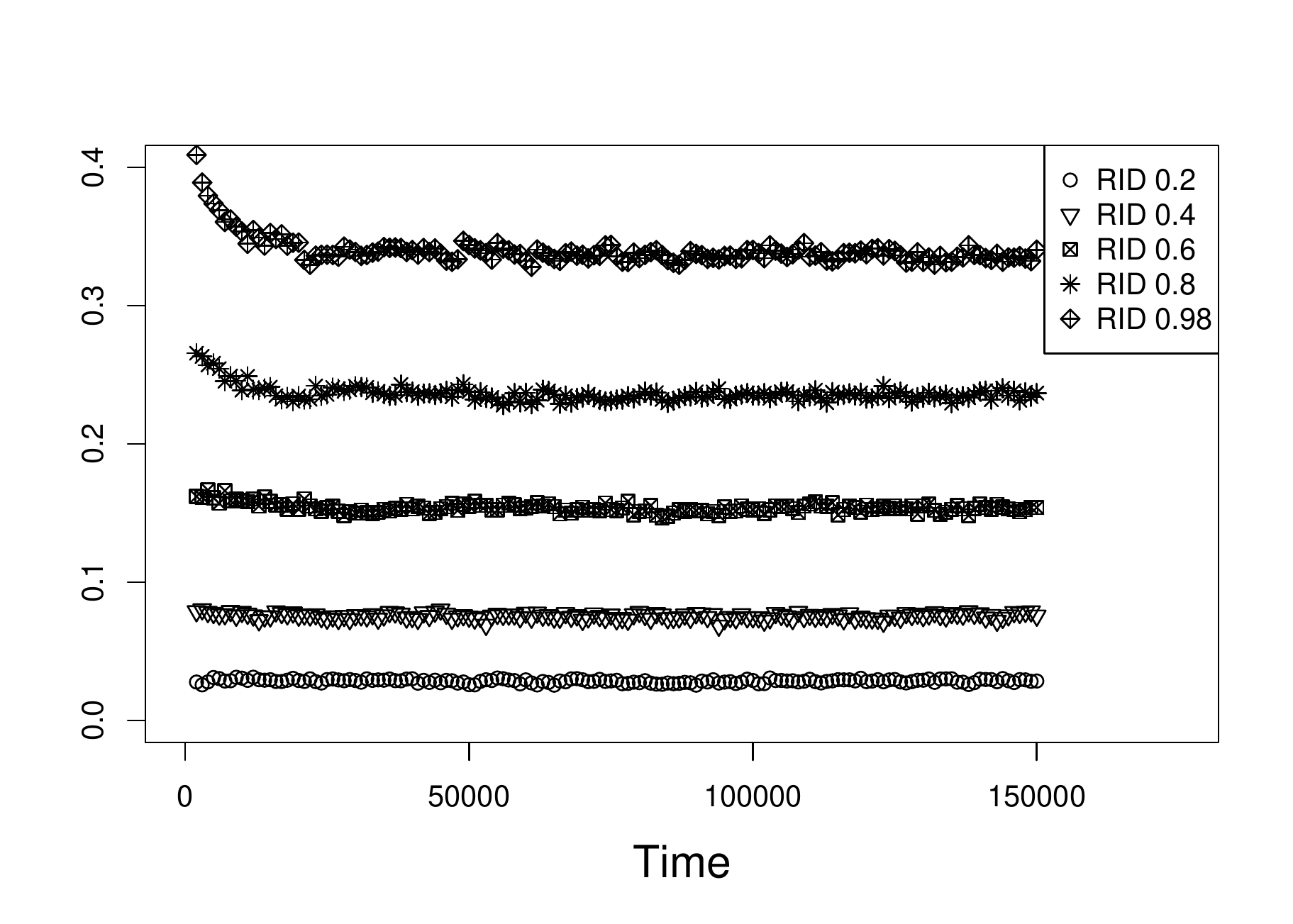}
            \caption{{\small $\bar{R}_w$ for AID Sario}}    
     \end{subfigure}
       \begin{subfigure}[b]{0.45\textwidth}   
            \centering 
            \includegraphics[width=\textwidth]{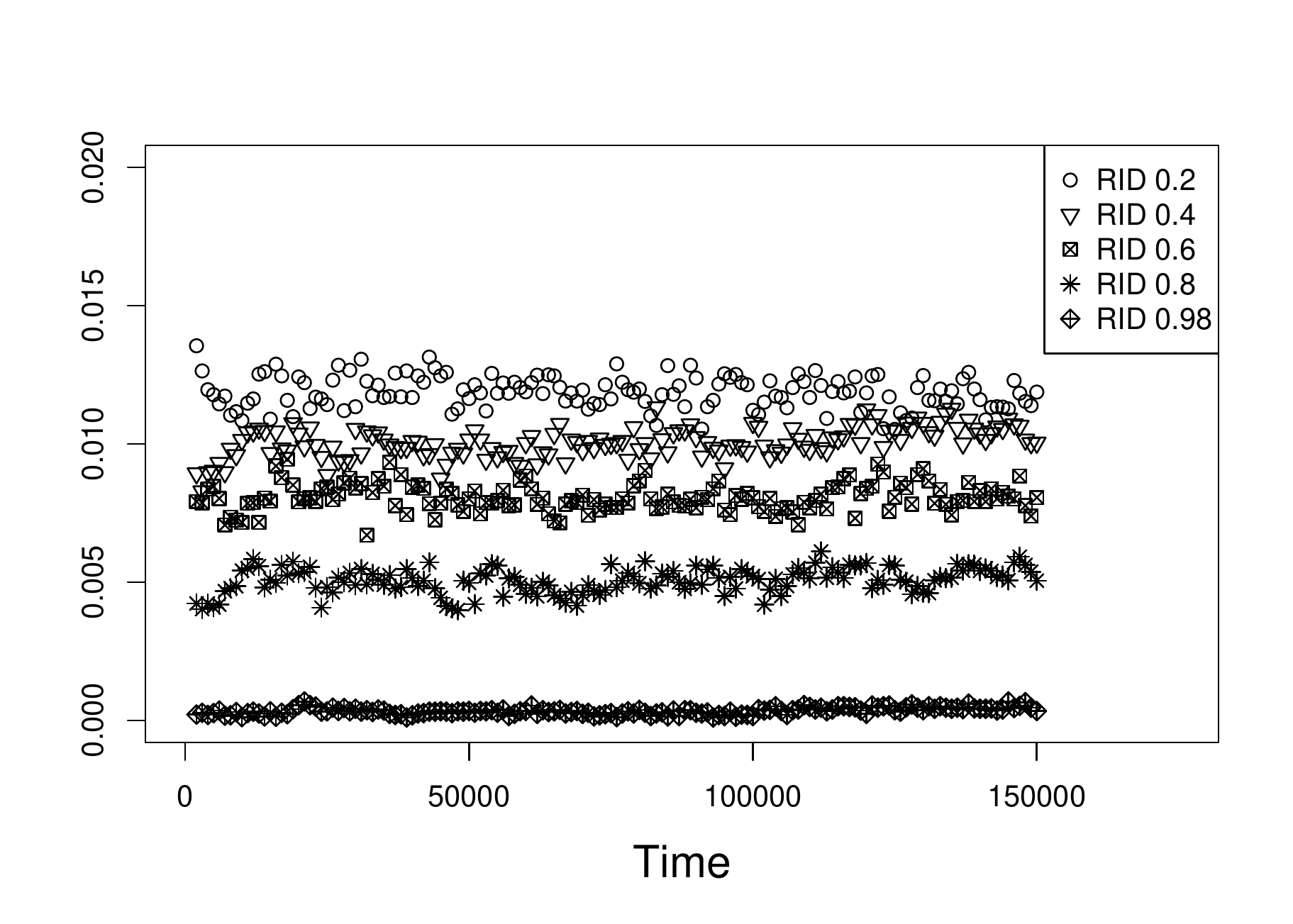}
            \caption{{\small $\bar{R}_b$ for AID Scenario}}    
        \end{subfigure}
        \caption
        {\small \textbf{Different average contact rates for RID and AID Scenarios:} For both RID and AID scenarios the between-group average contact rate  decreases, and the within-group average contact rate  increases as spatial fidelity increases.} 
        \label{fig:acr}
    \end{figure}

In both RID and AID scenarios, we also find as the spatial fidelity increases, the between-group average contact rate $\bar{R}_{b}$ decreases, and the within-group average contact rate $\bar{R}_{w}$ increases, Figure \ref{fig:acr}. Moreover, in the RM scenario, $\bar{R}_{b}$ and $\bar{R}_{w}$ are not significantly different (\textit{t=0.000108,P$>$0.99}).\\

\subsection{Information spread  dynamic I(t)}
In order to understand  how the spreading agents such as information or pathogen propagate over the contacting space, we   track the fraction of informed workers $I(t)$ in colony under different spatial fidelity and environment scenarios. 

\begin{figure}[!ht]
\centering
\begin{subsubcaption}
\begin{subfigure}{3.5cm}
\includegraphics[width=\linewidth]{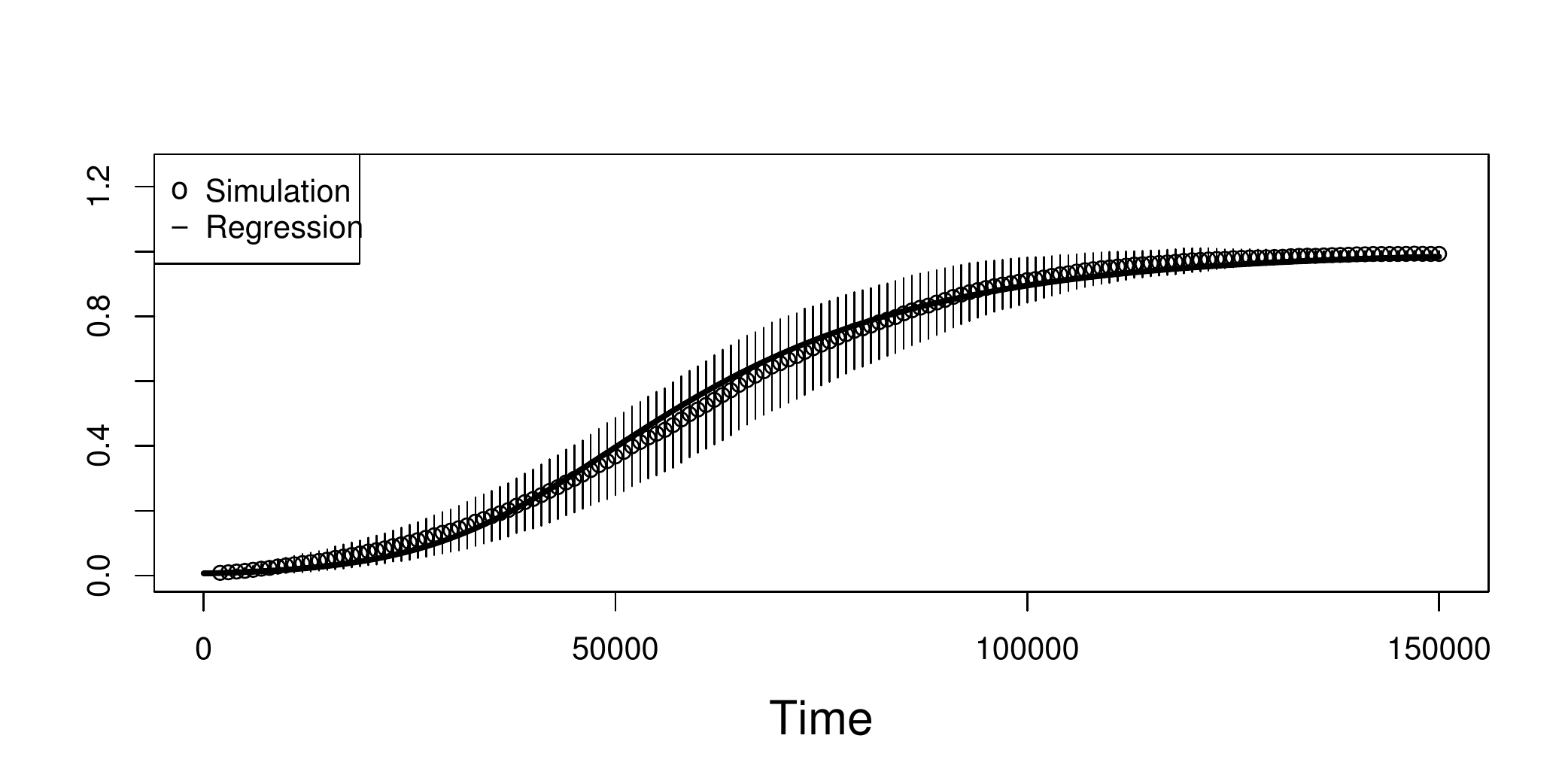}
\caption{RM with $SF=0$}
\end{subfigure}\qquad
\begin{subfigure}{3.5cm}
\includegraphics[width=\linewidth]{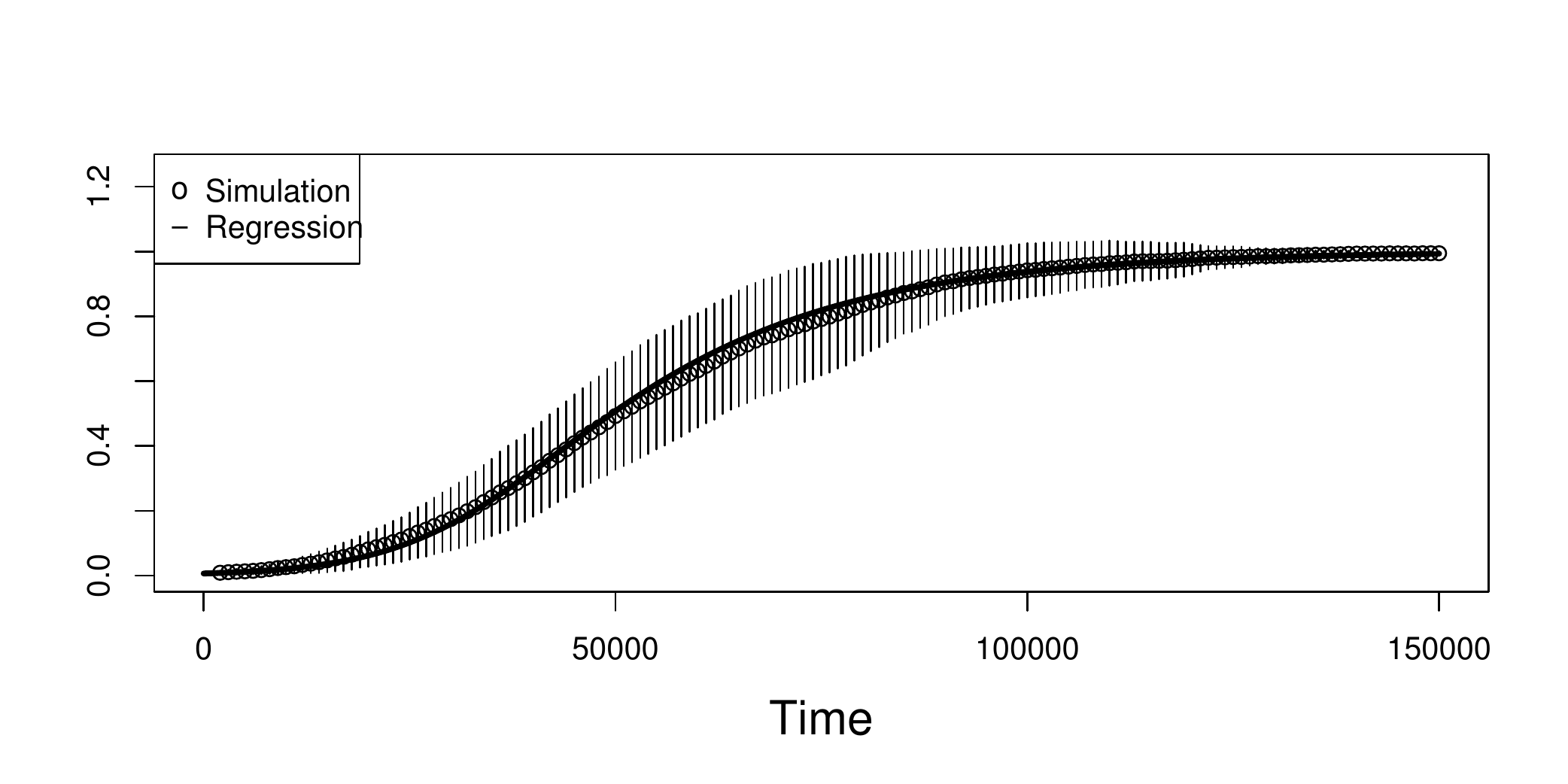}
\caption{RID with $SF=20\%$}
\end{subfigure}
\begin{subfigure}{3.5cm}
\includegraphics[width=\linewidth]{RID0Information}
\caption{RM with $SF=0$}
\end{subfigure}
\begin{subfigure}{3.5cm}
\includegraphics[width=\linewidth]{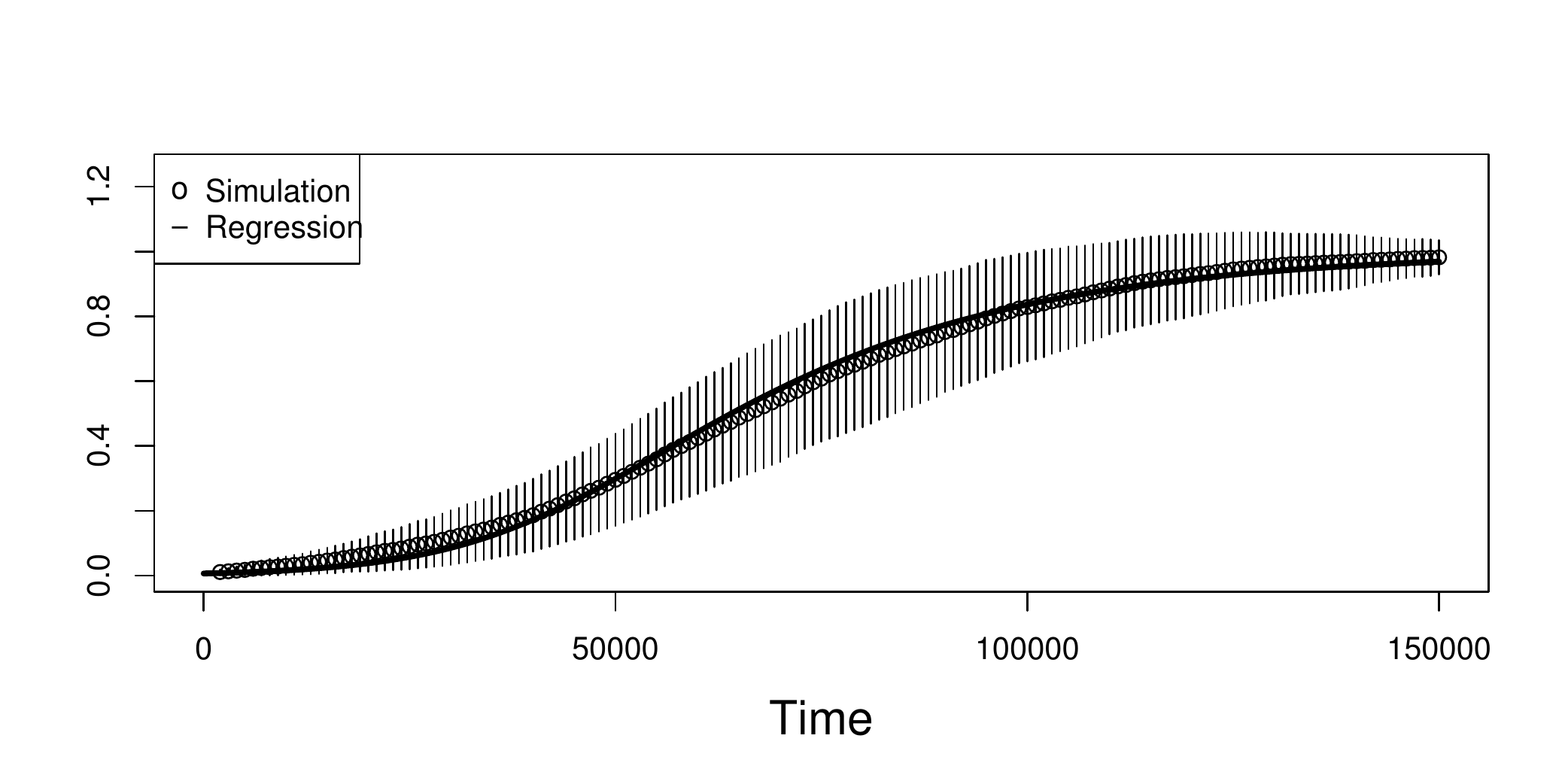}
\caption{AID with $SF=20\%$}
\end{subfigure}
\end{subsubcaption}

\begin{subsubcaption}
\begin{subfigure}{3.5cm}
\includegraphics[width=\linewidth]{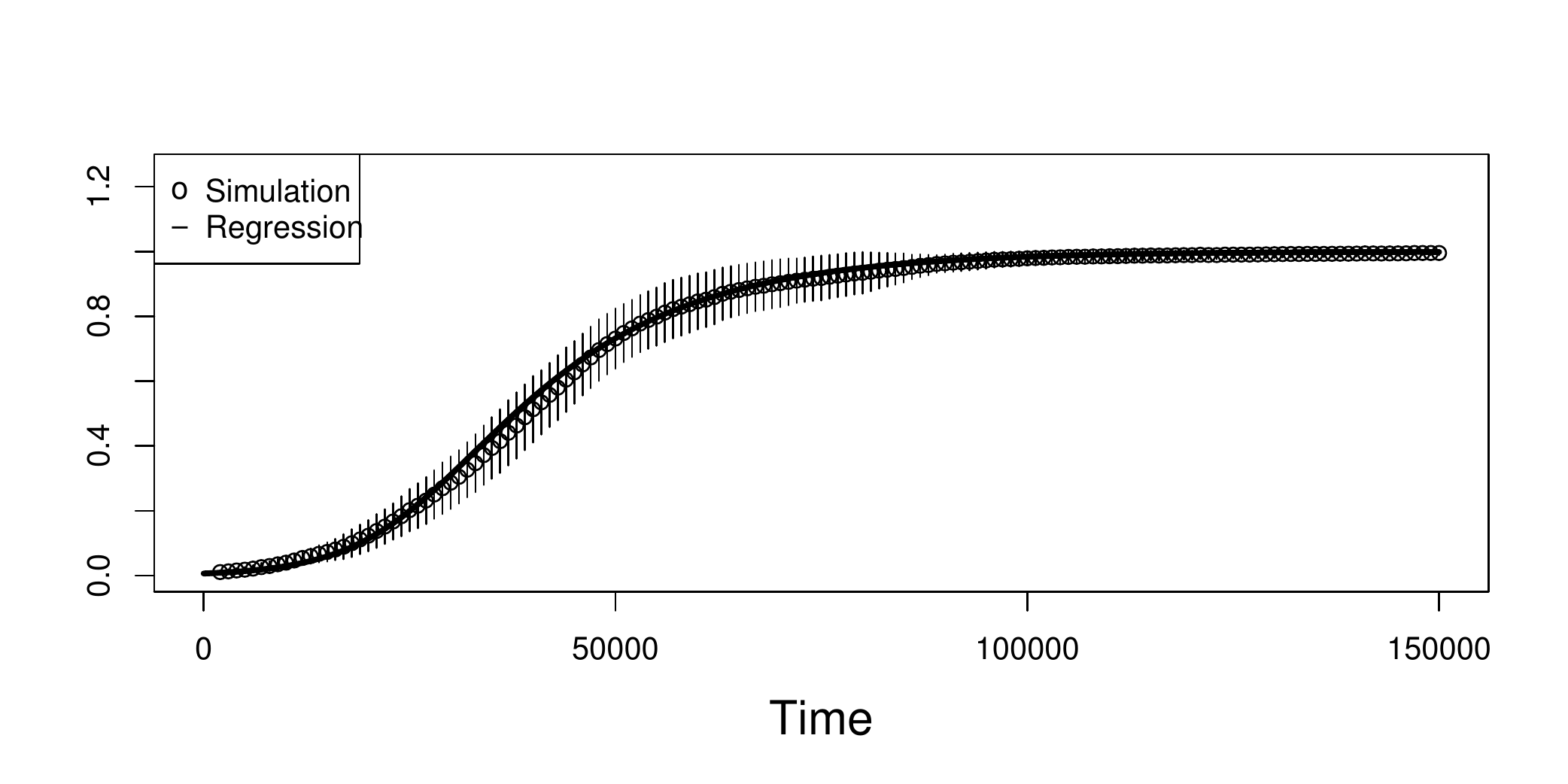}
\caption{RID with $SF=40\%$}
\end{subfigure}\qquad
\begin{subfigure}{3.5cm}
\includegraphics[width=\linewidth]{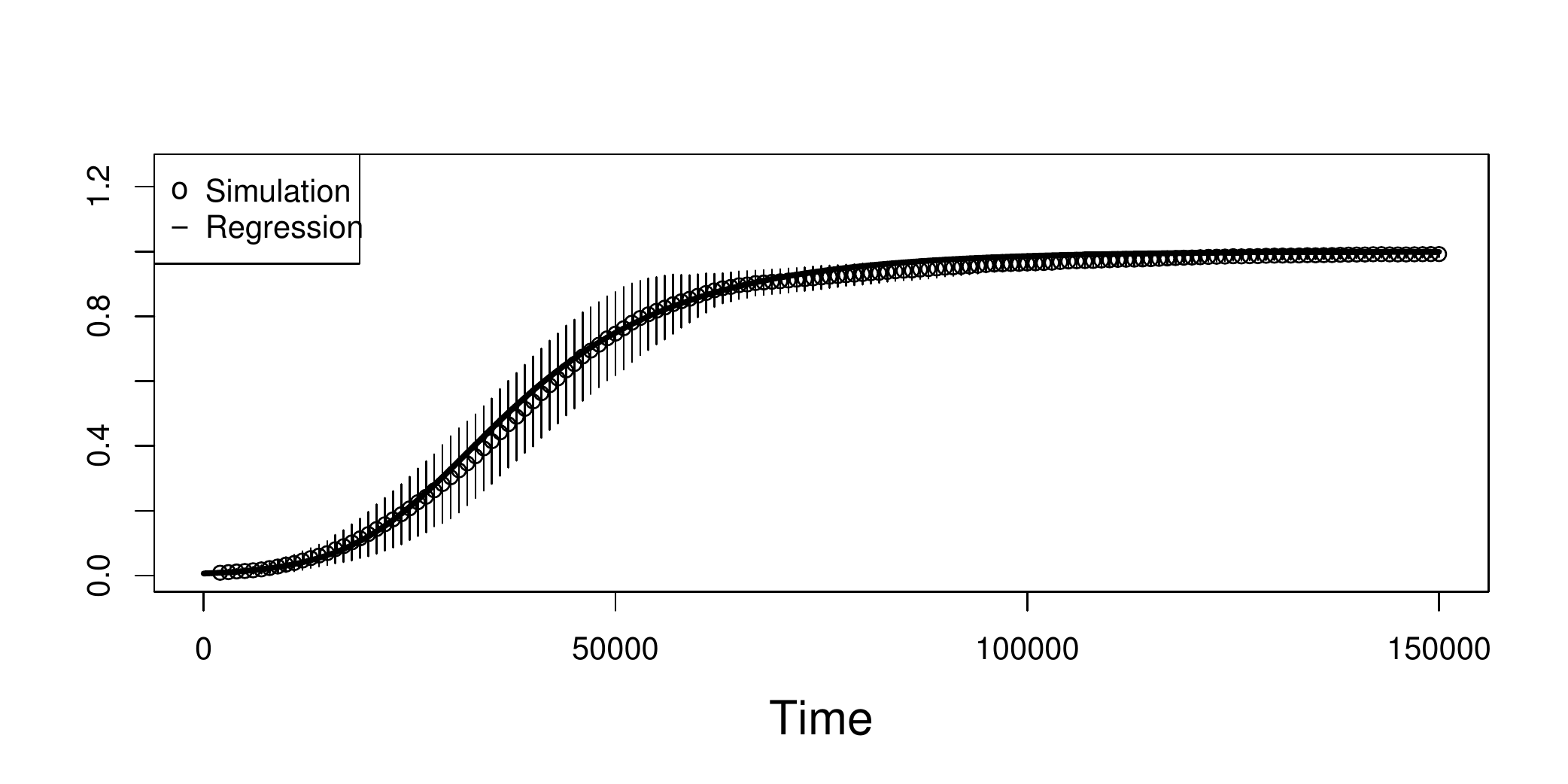}
\caption{RID with $SF=60\%$}
\end{subfigure}
\begin{subfigure}{3.5cm}
\includegraphics[width=\linewidth]{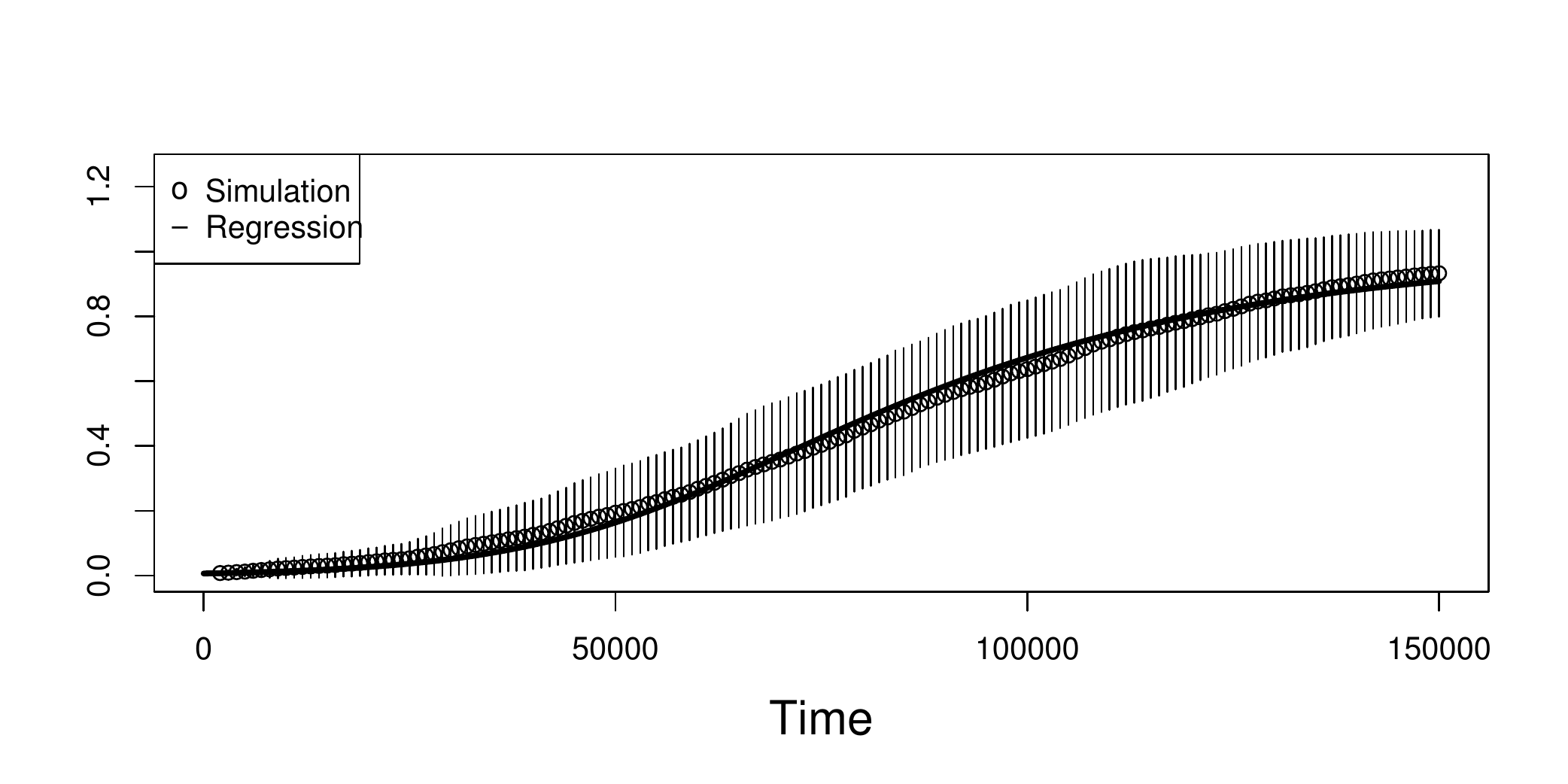}
\caption{AID with $SF=40\%$}
\end{subfigure}
\begin{subfigure}{3.5cm}
\includegraphics[width=\linewidth]{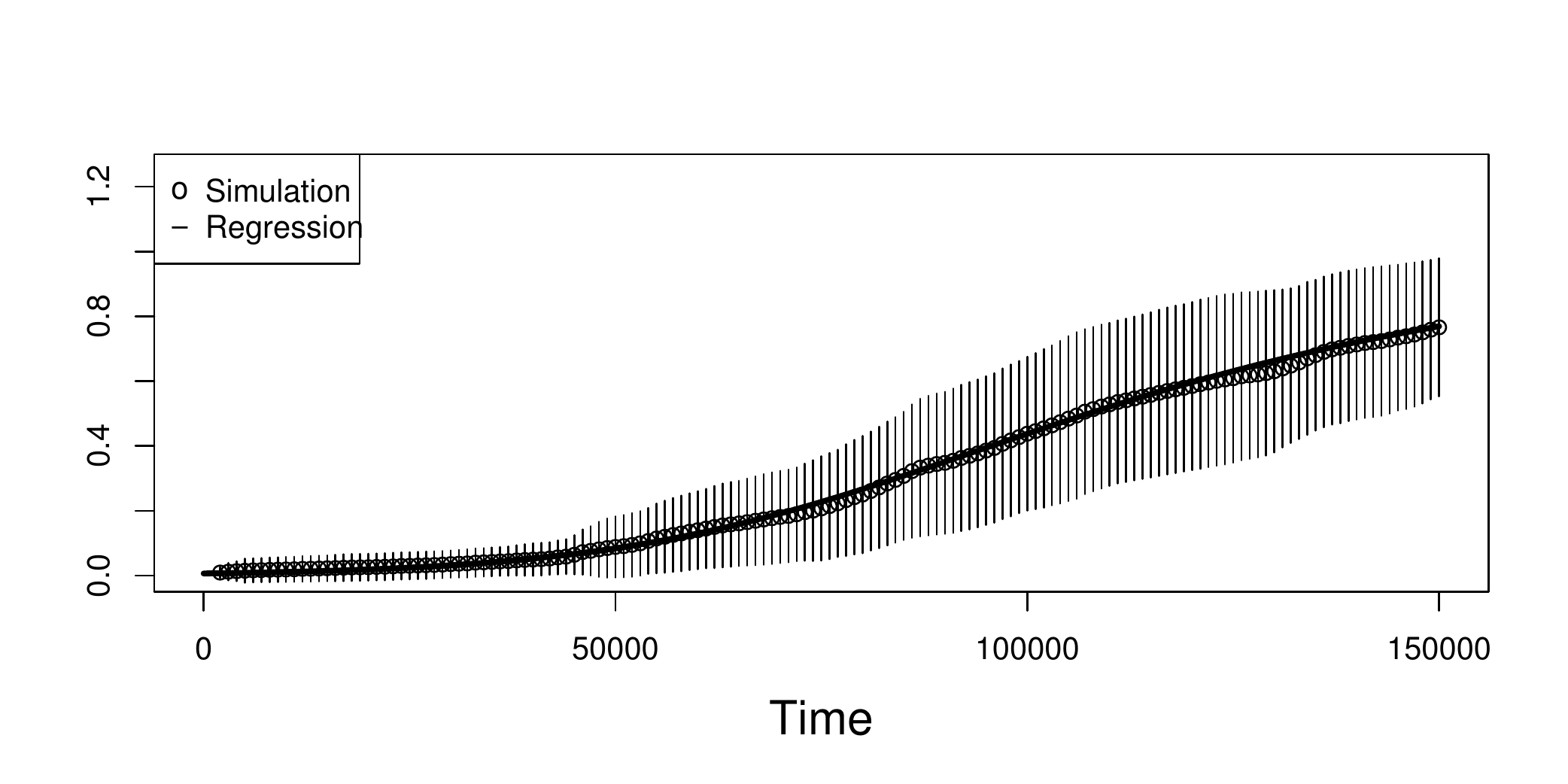}
\caption{AID with $SF=60\%$}
\end{subfigure}
\end{subsubcaption}

\begin{subsubcaption}
\begin{subfigure}{3.5cm}
\includegraphics[width=\linewidth]{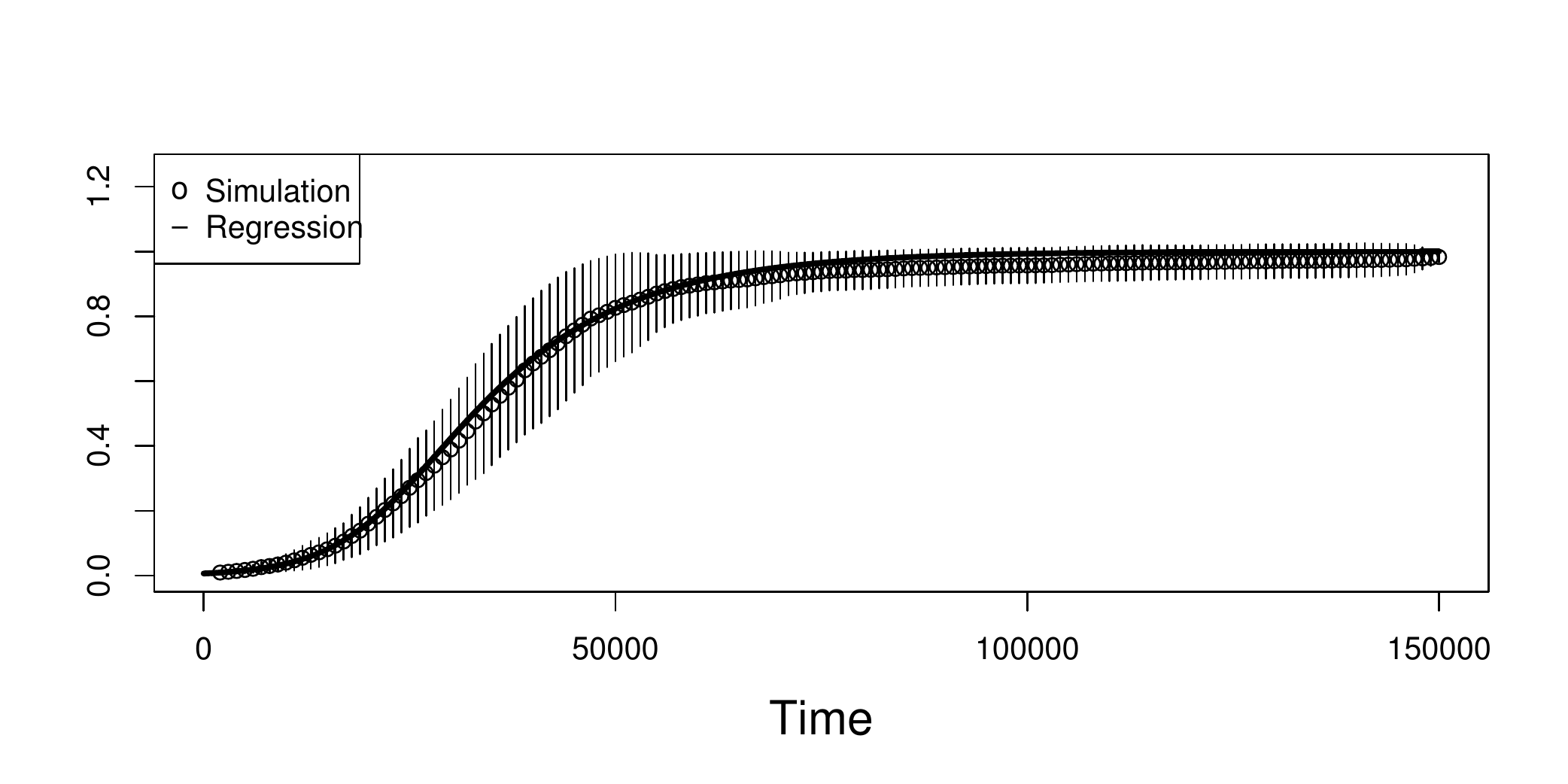}
\caption{RID with $SF=80\%$}
\end{subfigure}\qquad
\begin{subfigure}{3.5cm}
\includegraphics[width=\linewidth]{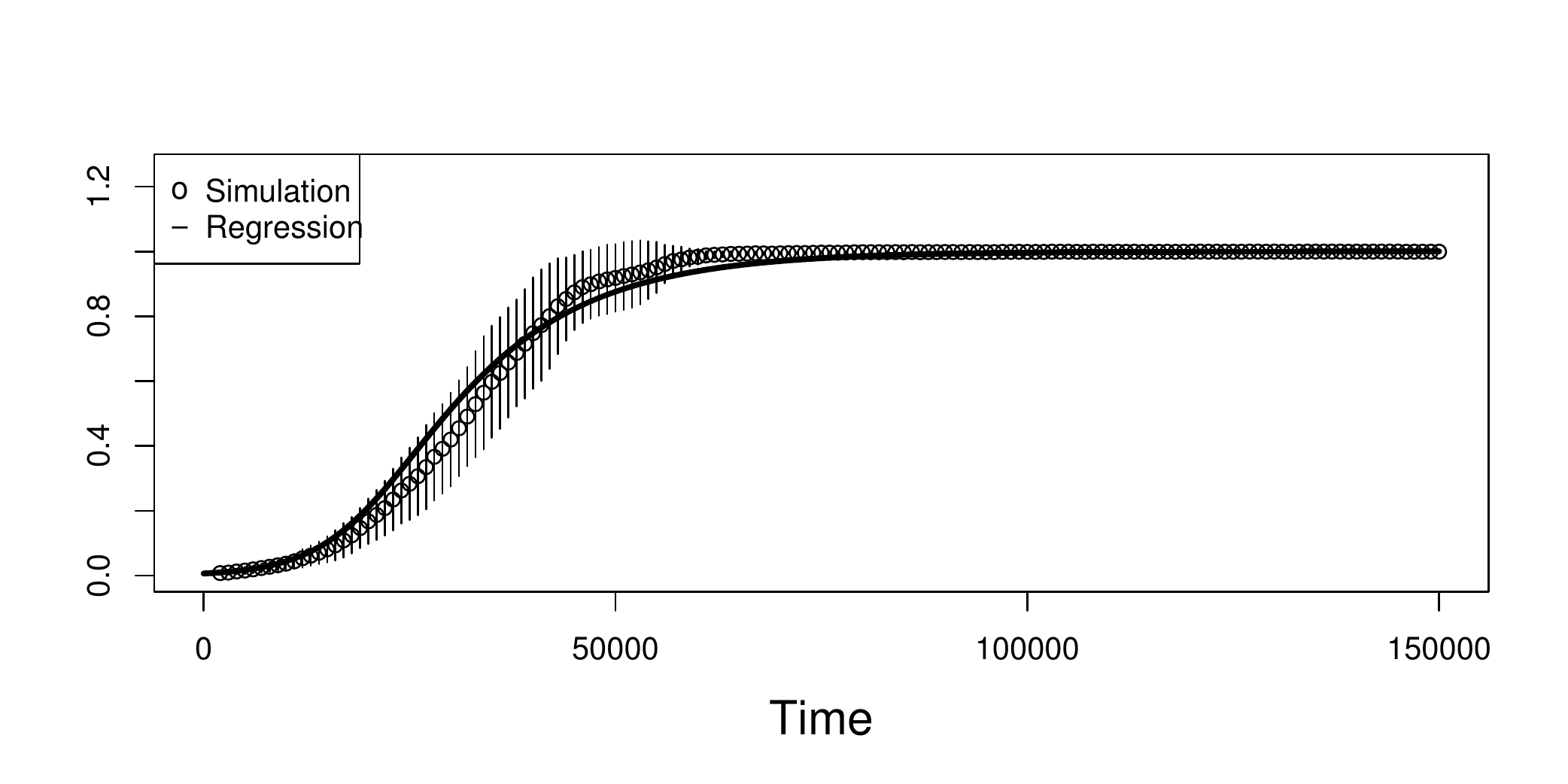}
\caption{RID with $SF=98\%$}
\end{subfigure}
\begin{subfigure}{3.5cm}
\includegraphics[width=\linewidth]{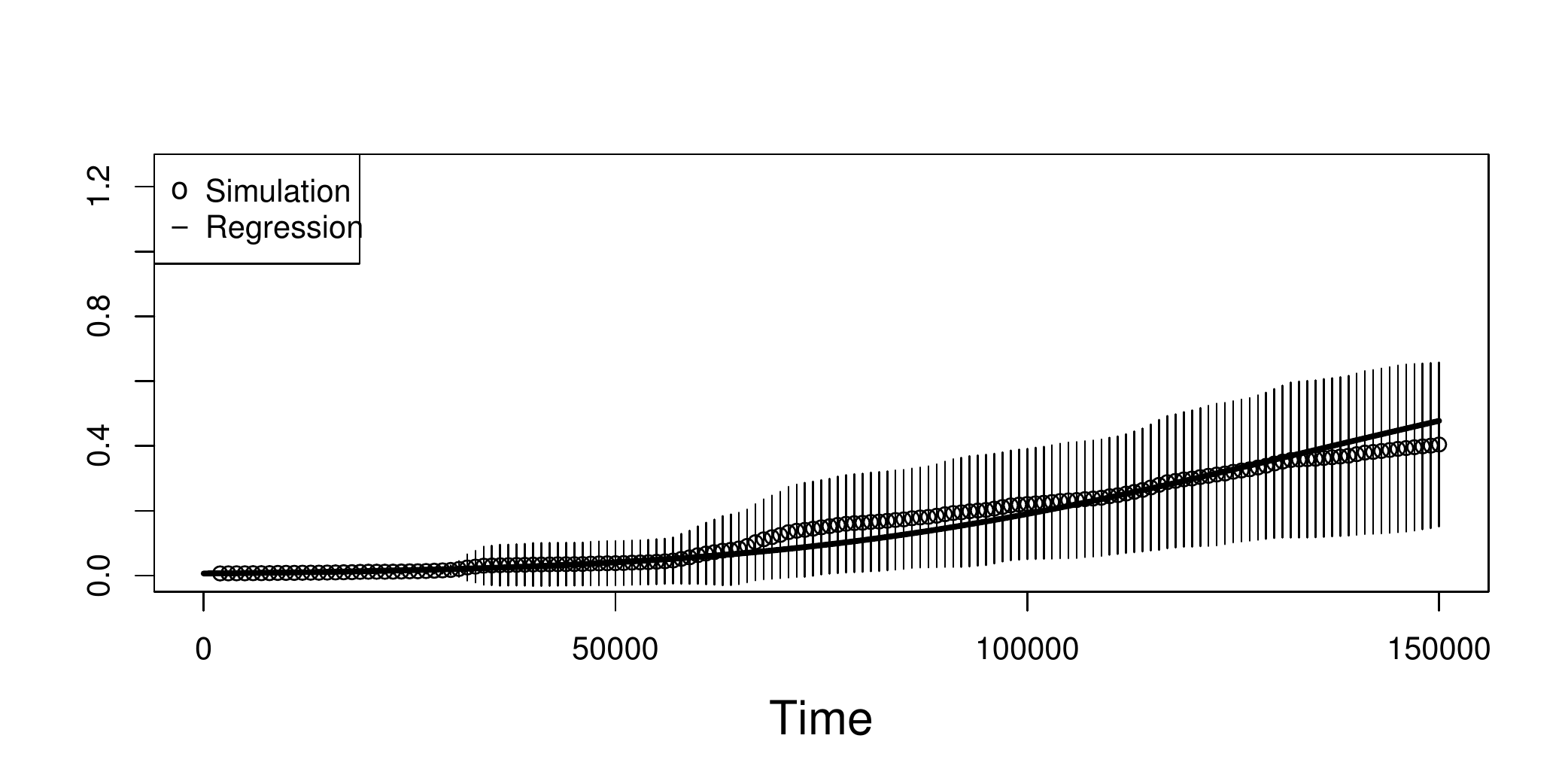}
\caption{AID with $SF=80\%$}
\end{subfigure}
\begin{subfigure}{3.5cm}
\includegraphics[width=\linewidth]{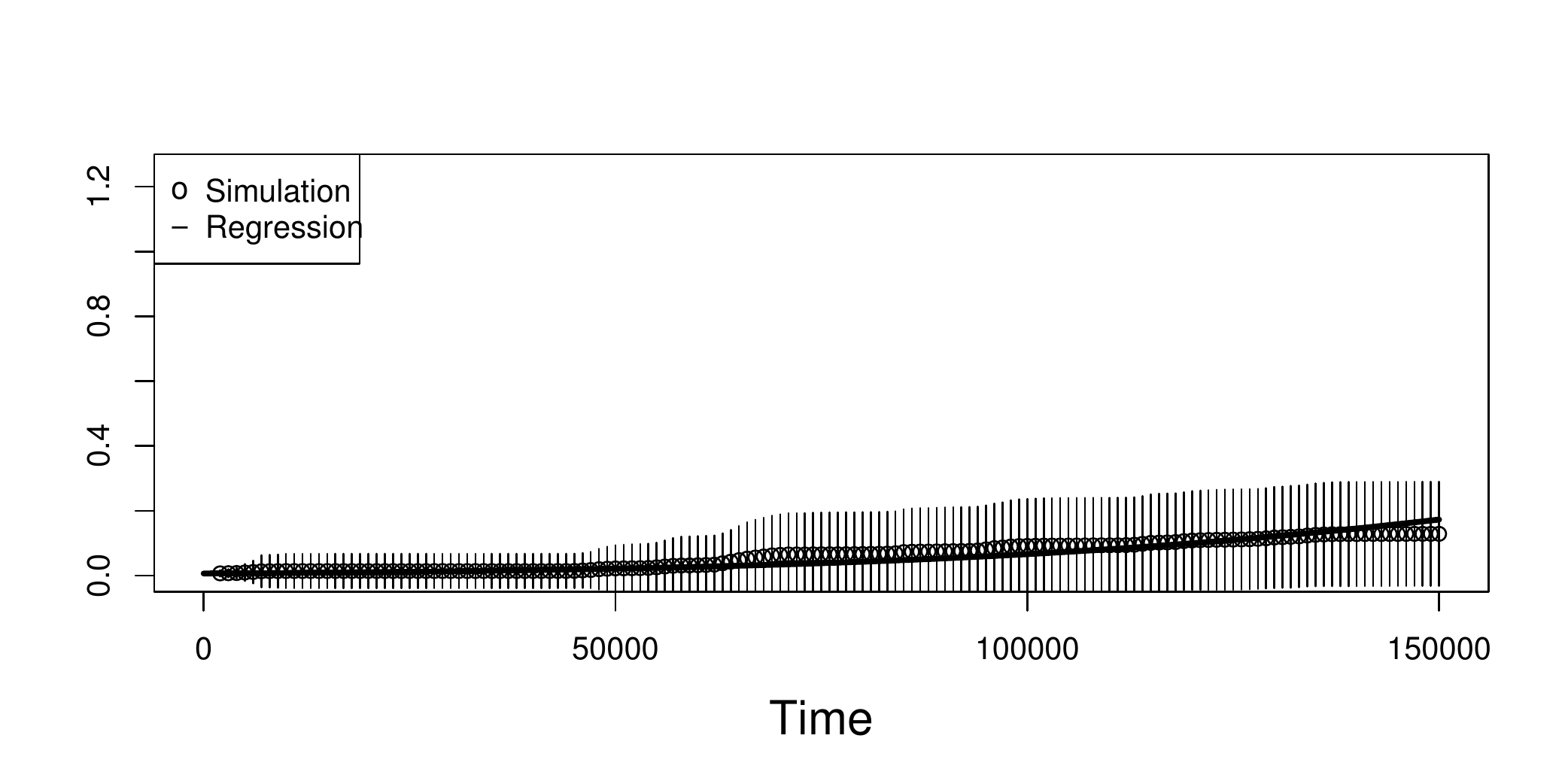}
\caption{AID with $SF=98\%$}
\end{subfigure}
\end{subsubcaption}

\caption{\label{fig:regression2}\normalsize{\textbf{Fraction of informed workers over time  in RM, RID, and AID scenarios and different $SF$ values:} The points represent the dynamic of the average of $40$ replicates and error bars are $95\%$ confidence intervals. The black curves are predicted values from the fitting model Equation \ref{equ6}. For RM and RID scenarios all the workers become informed at quasi-stationary state, however for AID
scenario the quasi-stationary state value depends on spatial fidality $SF$. }}
\end{figure}

For the RM and RID scenarios, the quasi-stationary state for the average fraction of informed workers is almost $100\%$, but for AID scenario, an outstandingly varied fractional workers ($12\%-97\%$ ) being informed in the end suggests that the inhibition of agents' transmission is probably caused by the spatial segregation, Figure \ref{fig:regression2}.\\

\begin{figure}[htp]
\centering
 \begin{subfigure}{0.45\textwidth}
        \centering
\includegraphics[width=\textwidth]{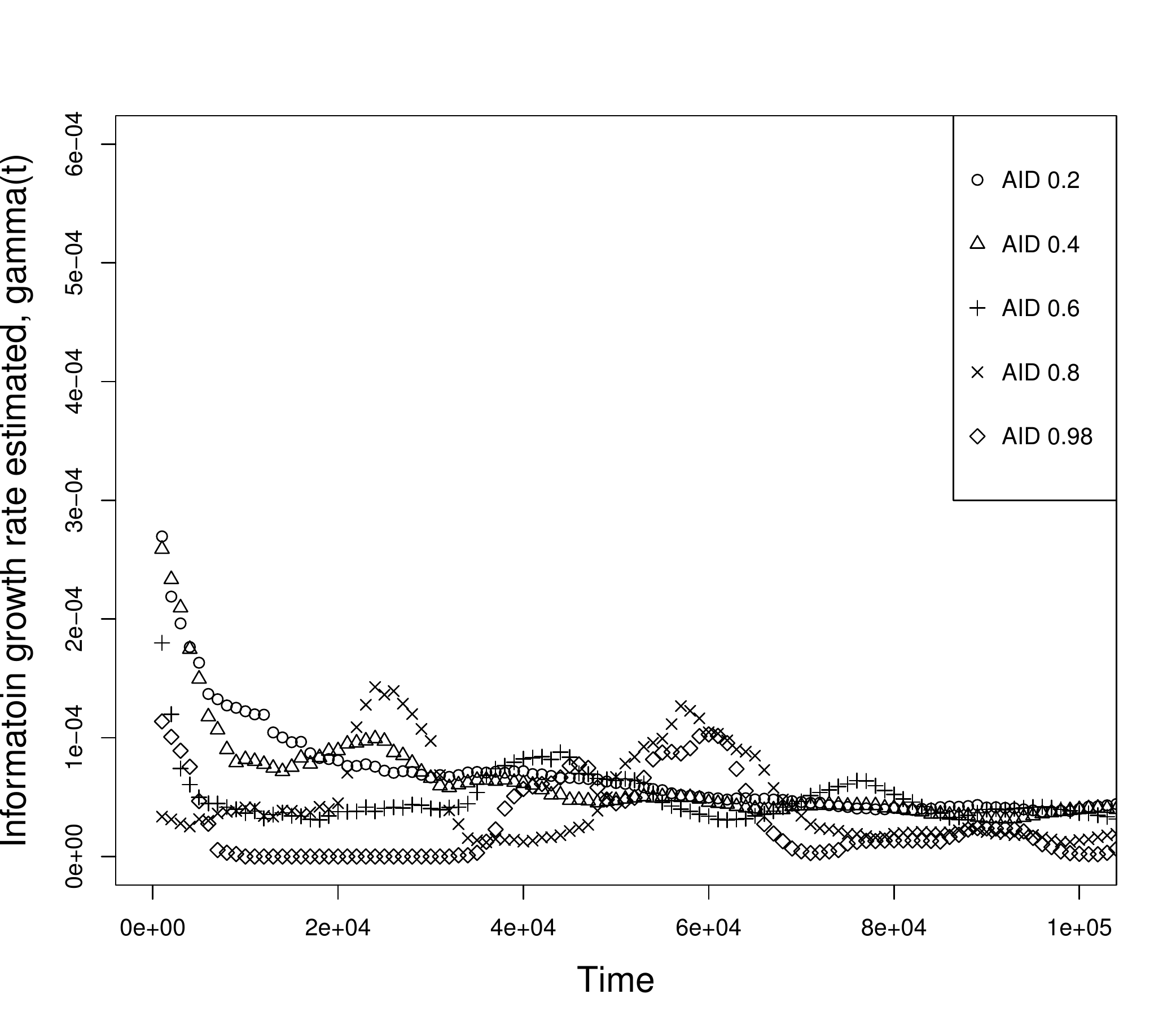}
\end{subfigure}
 \begin{subfigure}{0.45\textwidth}
        \centering
\includegraphics[width=\textwidth]{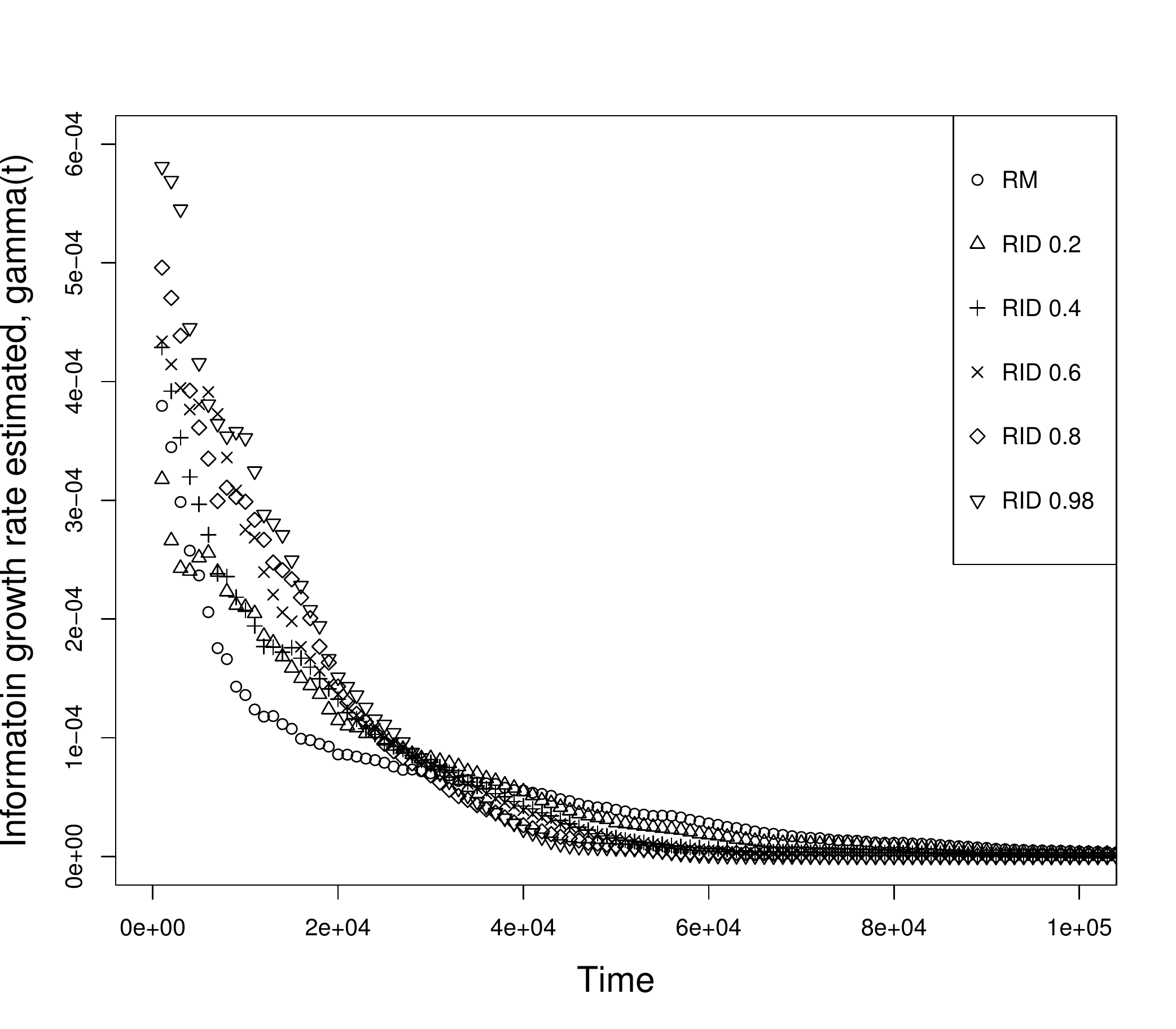}
\end{subfigure}
\caption{\label{fig:estimation} \normalsize{\textbf{Calculated \pmb{$\gamma(t)$}  over time in scenarios RM and RID (left panel) and  AID (right panel):}  $\gamma(t)$ decays exponentially over time in all scenarios.
}}
\end{figure}

Another important observation from dynamic of $I(t)$ is that spreading agents  follow a modified logistic growth pattern, Figure \ref{fig:regression2}. To identify how the  dynamics of agents correlates to the traditional non-spatial logistic growth model, we estimated the intrinsic growth rate  $\gamma(t)$ by using the following  equation
$$
\gamma(t)=\frac{I({t + \Delta t}) - I(t)}{\Delta t \cdot I(t) \cdot (1 - I(t))},$$
where $I(t)$ is the fraction of informed workers at time $t$ and $\Delta t=10000 $ is the time interval. The intrinsic growth rate $\gamma(t)$ decays over time in all scenarios, which is different from the constant rate in traditional non-spatial logistic model without space, Figure \ref{fig:estimation}.  The work on the effects of spatial correlation between the susceptibles and infected  by \cite{keeling1999effects} indicates that transmissibility of pathogens could be restricted by the identity of neighbor nodes in the  network. Thus we speculated the intrinsic growth rate $\gamma(t)$ in our spatial model can be a function of $e^{- I}$ to reflect the local saturation of transmission due to the restricted spatial connection between informed and non-informed workers.

Therefore, we perform the nonlinear regression by using the following modified-logistic model 

\begin{eqnarray}
\label{equ6}
\frac{{dI}}{{dt}} =& {R_{RM}} \cdot {Q_p} \cdot {e^{ - I(t)}} \cdot I(t) \cdot (1 - I(t)),
\end{eqnarray} 
where the carrying capacity of fractional informed workers equal to $100\%$, $R_{RM}=0.035$ is parameter of contact rate without spatial effect, which is the same for all scenarios and is  estimated from average of contact rate over time in RM scenario, and the parameter $Q_p$ is the transmission rate of spreading agents estimated from Equation \ref{equ6}, and $e^{-I(t)}$ is a encountering probability between informed and non-informed workers in the Poisson process. 
We estimated $Q_p$ in different scenarios and  different spatial fidelity $SF$ to examine effects of $SHD$, Table \ref{tab:q_estimate}. This estimation for $Q_p$ shows that in RID scenarios the larger $SF$ gives larger $SHD$ and larger $Q_p$, but in AID scenarios larger $SF$ gives smaller $Q_p$. Our regression model fits the sigmoid curves of agents' spread robustly for all scenarios, Figure \ref{fig:regression2}. We also observed that the larger value of $Q_p$, faster the fractional informed ants arrive to the plateau. 

\begin{table}[H]
 \centering 
   \begin{tabular}{lll}
   \toprule[\heavyrulewidth]\toprule[\heavyrulewidth]
   \textbf{ \pmb{$SF$}} & \textbf{ \pmb{$Q_p$} for  AID} & \textbf{\pmb{$Q_p$} for RID}  \\ 
   \midrule
   $20\%$ & $0.002457$***& $0.00331$***\\
   $40\%$ &$0.002014$***& $0.00417$***\\
   $60\%$&$0.001526$***&$0.00443$***\\
   $80\%$&$0.001083$***&$0.00491$***\\
   $98\%$&$0.000691$***&$0.00563$***\\
   \bottomrule[\heavyrulewidth] 
   \end{tabular}
   \caption{\normalsize{$Q_p$ estimation  for Equation \ref{equ6}, the asterisk indicates statistical significance.}}
   \label{tab:q_estimate}
\end{table}

\begin{figure}[htp]
\centering
\includegraphics[width=0.7\textwidth]{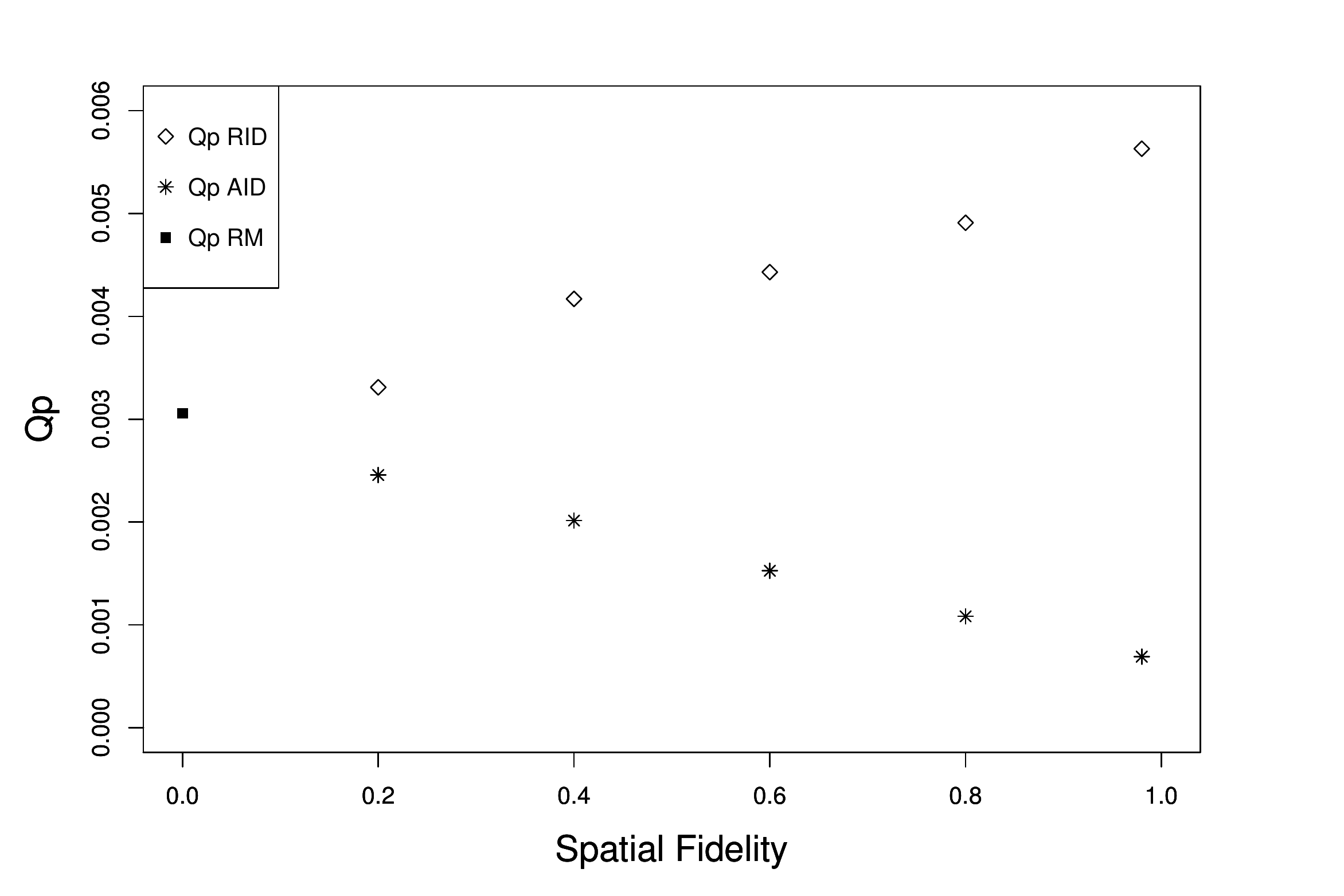}
\caption{\label{fig:bifurcation}\normalsize{\textbf{Pairwise comparisons of transmission rate for RM, RID, and AID  scenarios: } There is a positive linear correlation between transmission rate $Q_p$ and spatial fidelity $SF$ for RID scenario indicating its transmission promotion effects, but the correlation between $Q_p$ and $SF$ for AID scenario  is   negative because of  transmission inhibition effects.}}
\end{figure}

To further study the correlation between $Q_p$ and $SF$ values under different scenarios, we plot the spatial fidelity $SF$ versus $Q_p$, Figure  \ref{fig:bifurcation}. The result provides a visual presentation: circles are RID scenarios, triangles are for AID scenarios, and the diamond is the RM scenario. We observe that there is a bifurcating pattern of $Q_p$ as a linear function of $SF$ in RM, RID, and AID scenarios. The linear fits for RID and AID are shown as follow:
\[ \begin{cases} 
     Q_p= 0.003 + 0.0026 SF, & \textnormal{RID scenario} \\
      Q_p=0.003 -0.0024 SF & \textnormal{AID scenario}
   \end{cases}
\]

\section{Discussion}

The flow of spreading agents within a biological social network is not random. Instead, heterogeneity among individuals in their communication clusters and in their spatial distributions  influences spreading agents  across groups. For a social insect colony, in which individual behavior depends on their task, both spatial and network heterogeneity are driven by individuals task roles at any given time. In this paper, we assigned task roles to individual agents and manipulated individual spatial preferences and initial conditions to explore the impact of spatial behavior on social contacts and agents' transmission. Our dynamical model included three different task groups, with their corresponding SFZs. We additionally studied the impact of task on movement, by assigning workers in social insect colonies with different tasks to either random or preferential walking styles. We will discuss dynamical effects on the processes of social contacts and agents spreading in the following four different aspects:\\

{\noindent\textbf{Dynamics of social contacts:}} The individual interactions in social insects colony were the straightforward pathway to inseminate and transmit spreading agents such as information or pathogens in a contacting network \cite{pacala1996effects}. Ant workers were found to change their contact rates flexibly over time to regulate local information capturing \cite{gordon1996organization,gordon1993function}, e.g. restricting the flow rate of spreading agents through regulating the contact rate in a time-order network \cite{blonder2011time}. In our model simulations, the probability of contact between workers depends on their neighboring space. Meanwhile, we observed that contact makes varied contributions on the spreading agents propagation. For example, for the scenarios with $40\%$ spatial fidelity, the functional contacts contributing to agents' spread were only accounted for less than $1\%$ of total contacts when initial aggregation of  workers, but $7\%$ when their initial distribution is random. It was suggested that the spatial correlation between the informed and non-informed  workers in the local scale might interfere with an expected speed of transmission of spreading agents \cite{keeling1999effects}. As the spatial fidelity escalates the spatial heterogeneity degree, information about tasks is more likely to be transmitted within groups in  colonies, which may be a potential mechanism to maintain the task specialties \cite{naug2009structure}. When the colony has extremely high spatial fidelity (e.g., $98\%$) with aggregated initial distribution (AID), the propagation of spreading agents highly relies on the contacts between groups through random walkers.   One of the consequences is that the high spatial fidelity results in the slower transmission rate of agent, e.g. pathogens, which is probably one of mechanisms of social immunity in the social insects colonies  \cite{cremer2007social}.\\

%
\noindent\textbf{Spatial effects:}
The fraction of informed workers in our simulation shows an obvious logistic-pattern which corroborates the finding of previous studies on mobile encounter networks \cite{adler1992information,arai1993information,korhonen2007logistic,pratt2005quorum}, a food trophallaxis network in an ant colony \cite{greenwald2015ant,sendova2010emergency} and contagious pathogen model simulations for social insect colonies \cite{naug2002role}. Comparing to the standard logistic growth model without spatial components, modifications in Equation \ref{equ6} imply that spatial effects, such as local spatial correlation, cluster distribution and preferential movement of  workers may distort the linkage between physical contagion and mass action of spreading agents. The modified-logistic model in Equation \ref{equ6} uncovers two main spatial effects: local saturation of spreading agents and spatial segregation of workers. In the correlation model \cite{keeling1999effects}, the local spatial correlation between the susceptible and the infected ones was found to lead the reproductive ratio of spreading agents to decay over time after the single infectious individual invades a cluster of susceptible individuals. \\

The other spatial effect that can be observed is the strong linear relation between agents spatial fidelity and the transmission rate of spreading agents, $Q_p$ in different scenarios, Figure \ref{fig:bifurcation}. In general, the estimates of transmission rate $Q_p$ in Figure \ref{fig:bifurcation} suggest the dual-functionalities of spatial fidelity on agents' transmission rate in scenarios. When the initial distribution of  workers is aggregated, the structure of spatial clusters induced by workers preferential movement heterogenized the neighboring space of the non-informed/informed  workers, and shielded  workers from being exposed to external spreading agents. The inhibiting effects of spatial fidelities on spreading agents are similar as the clustering effects that restrict the potential further transmission across household \cite{grassly2008mathematical}. Specifically, the biological barriers in colonies arising from spatial aggregations are one of the mechanisms of organizational immunity \cite{feigenbaum2010influence,naug2007experimentally}. Meanwhile, we observed the promoting effect of spatial fidelities on spreading agents in random mixing and random initial distribution scenarios, Figure \ref{fig:bifurcation}. Intuitively, workers directional movements arising from the initial random positions would intensify the mixing effects and help agents being transmitted over the colony. Spreading agents, such as food have been observed to spread faster and more uniformly in the groups with better spatial mixing among individuals in the colonies of honeybee \textit{Apis mellifera} \cite{naug2008structure} and the ant \textit{Temnothorax albipennis} \cite{sendova2010emergency}\\

\begin{figure}[!ht]
\centering
\includegraphics[width=0.7\textwidth]{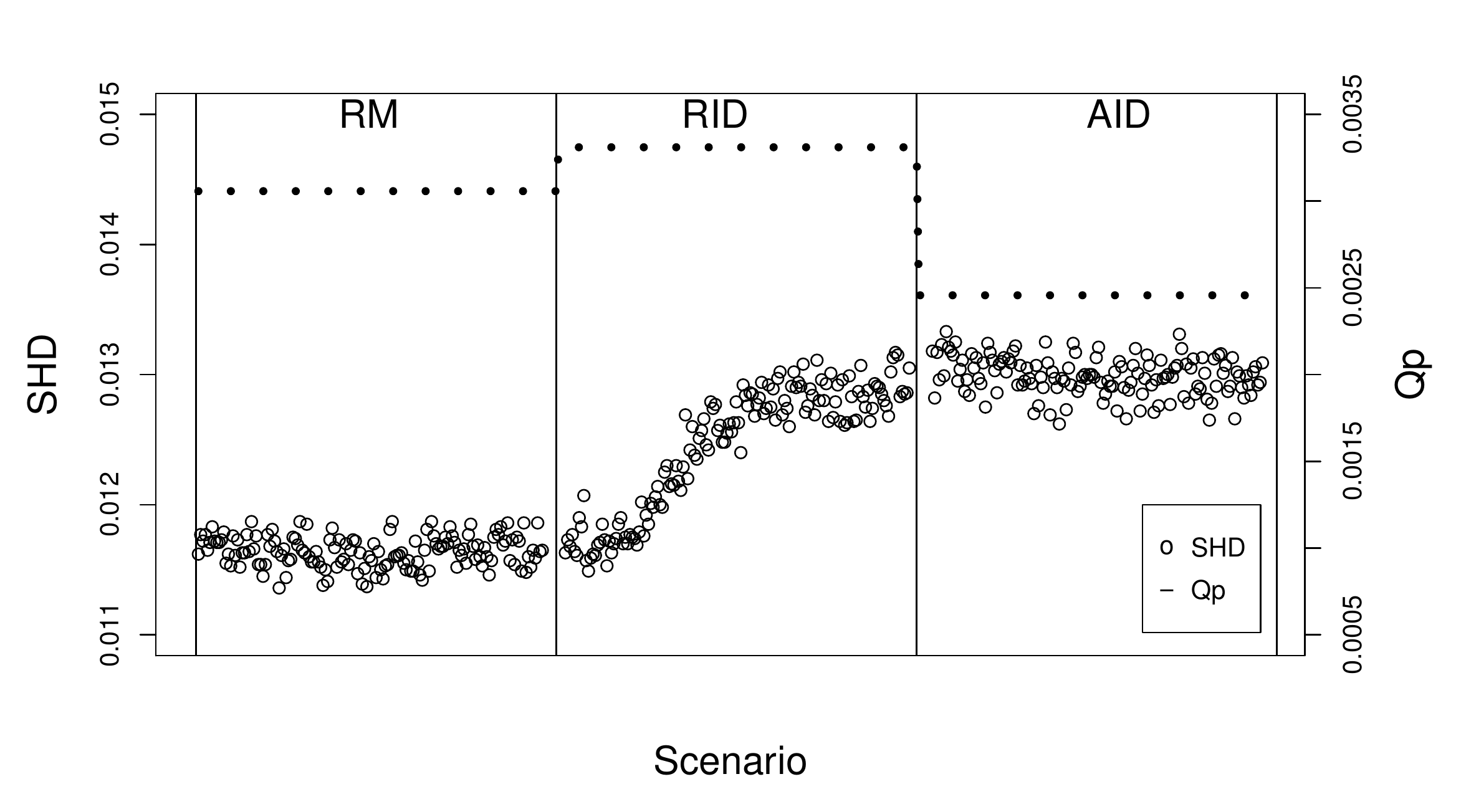}
\caption{\label{fig:dual} \normalsize{\textbf{Possible adaptive spatial strategies in three scenarios with \pmb{$SF=20\%$}:} Social insect colonies can regulate the  cost and benefits arising from properties of spreading agents during the consecutive scenarios, RM$\rightarrow$RID$\rightarrow$AID.}}
\end{figure}

\noindent\textbf{Environmental effects on spreading agents:}
The trade-off between beneficial and harmful spreading agents through social insect colonies could be resolved by mechanisms of encountering networks and diffusion of chemical signals \cite{blonder2011time,pinter2015persistent,richardson2015beyond,sendova2010emergency}. Our model simulation provides an alternative explanation for the trade-off through changes in individual spatial behavior induced by environmental events/scenarios. In social insects colonies, the spatial distribution of  workers has been observed to change in response to environmental events instantly. For example, under threats, workers break down their spatial tendency and mix randomly as an effective strategy to relieve threats \cite{sendova2010emergency,wilson1971evolution}. Without immediate threats, it was found that spatial segregation provided colonies protection against pathogens exposure \cite{fefferman2007disease,naug2002role,pie2004nest,stroeymeyt2014organisational}. Thus, the opposite effects of workers' spatial behavior on transmissions of spreading agents demonstrates the capability of social insect colonies to regulate cost and benefits arising from properties of spreading agents during the consecutive scenarios, RM$\rightarrow$RID$\rightarrow$AID, Figure \ref{fig:dual}.\\

\noindent\textbf{Significance of spatial behavior:} 
Individual movement patterns heterogenize the probability of being exposed to spreading agents \cite{stroeymeyt2014organisational}. Pinter-Wollman \citep{pinter2015persistent} suggested that  workers' persistence in walking orientation may facilitate the information flow in a restricted space due to high interaction rates. We found the same effects of spatial fidelities in the RID scenarios: high spatial fidelity of  workers on each task group leads high contact rate of workers, and as consequences, speeds up transmission of spreading agents when the proportion of the workers (i.e., ones with the preferential walking style) persist in orientation and walk to SFZs from initial random positions.  High spatial fidelities could maximize the benefit of the agent transmission rate $Q_p$ in the environment with threat, and minimize the cost of $Q_p$ in normal environments with pathogens. Nevertheless, workers in the colonies of \textit{T. rugatulus} and \textit{Leptothorax longispinosus} were found to spend non-negligible amounts of time on wandering in the nest \cite{charbonneau2015lazy,charbonneau2015workers,cole1986social}. We speculated it would be beneficial to maintain some proportion of random walkers as a way to ensure the instant responses to local threatening events in the transitional scenario from initial spatial segregation to random mixing (AID$\rightarrow$RID). \\

Social insect colonies are an excellent example of complex adaptive systems, whose inter-individual interactions at local scales facilitate information spreading or inhibit pathogen transmission at global scales. Spatial heterogeneity generated by variations in individual task roles affect social contact dynamics, and thus the way in which agent spreads through social networks. We use variations in movement patterns associated with different tasks to build and study an agent-based model of social contact dynamics and the related agent spreading dynamics. Our proposed model incorporates the following three components that generate spatial heterogeneity: 1) three task groups, each assigned a general spatial zone in which the task is preferentially conducted; 2) variations in initial distributions of individuals, from general (random) mixing to aggregated one; 3) variations in working style associated with task roles, modeled either as a random walk, or via bias in turning radius towards the task zone. In this study, we found the spatial fidelity of social insects associated with task allocation and environmental events is the ultimate reason for variable transmission rates of spreading agents under the different conditions. We showed individual spatial/task fidelity is able to induce the task aggregation structure that has double-effects: 1) highly inhibiting the opportunity of being exposed to the external stimuli with initial aggregation scenario; 2) facilitating the encountering and agents exchanging with initial random distribution scenario. Those findings can help us understand the function of flexibility of social insects behavior under a changing environment.\\

In our future work, we will more focus on task switching in RID scenario to study how the social insects employ their spatial behavior to regulate information flow with a limited transmissibility, e.g. task cues rather than alarm signal. We are building a model based on attenation-networks with several mechanisms, e.g. individuals have spatial preferences based on spatial fidelity and mission location density, and individuals could switch their tasks based on the task cues captured from their neighbors. Also, those simulation results inspired us to conduct experiments to track how spatial clusters of social insects affect the information flows, e.g. alarm signal propagation in the colony.\\

\section*{Acknowledgments}
This research is partially supported by NSF-DMS (Award Number 1313312 \& 1716802);  NSF- IOS/DMS (Award Number 1558127); DARPA -SBIR 2016.2 SB162-005 Phase II; and The James S. McDonnell Foundation 21st Century Science Initiative in Studying Complex Systems Scholar Award (UHC Scholar Award 220020472).\\


\end{document}